\documentclass[useAMS,usenatbib]{mn2e}
\usepackage{epsfig} 
\usepackage{txfonts} 
\usepackage{multirow}
\usepackage{fmtcount}
\usepackage{graphicx}
\usepackage{enumerate}
\usepackage{epstopdf}

\newcommand{\be}{\begin{eqnarray}}
\newcommand{\ee}{\end{eqnarray}}

\newcommand{\rdec}{$R_{dec}$}

\newcommand{\go}{$\Gamma_0$}

\newcommand{\tpeak}{$t_{peak}$}
\newcommand{\adiab}{\hat{\gamma}}

\newcommand{\nr}{non--relativistic}
\newcommand{\fb}{fireball}
\newcommand{\ag}{afterglow}
\newcommand{\bw}{blast wave}

\newcommand{\epse}{$\epsilon_{e}$}

\newcommand{\epsp}{$\epsilon_{p}$}
\newcommand{\epsrad}{$\epsilon_{rad}$}
\newcommand{\lf}{Lorentz factor}
\newcommand{\eq}[1]{Eq.~\ref{eq:#1}}
\newcommand{\fig}[1]{Fig.~\ref{fig:#1}}

\title[Afterglow emission in GRBs]
{Afterglow emission in Gamma-Ray Bursts: I. Pair-enriched ambient medium and radiative blast waves}
\author[L. Nava et al.]
{L.~Nava,$^{1}$\thanks{E-mail: lara.nava@apc.univ-paris7.fr} 
L.~Sironi,$^{2}$\thanks{NASA Einstein Fellow} 
G.~Ghisellini,$^3$
A.~Celotti,$^{3,4}$
and G.~Ghirlanda$^3$\\
$^1$APC, Universit\'e Paris Diderot, CNRS/IN2P3, CEA/Irfu, Obs. de Paris, Sorbonne Paris Cit\'e, France\\
$^2$Harvard-Smithsonian Center for Astrophysics, 60 Garden Street, Cambridge, MA 02138, USA\\
$^3$INAF - Osservatorio Astronomico di Brera, via E. Bianchi 46, I-23807, Merate, Italy\\ 
$^4$SISSA - Via Bonomea 265, I-34136, Trieste, Italy\\
}
\begin{document}

\date{}

\pagerange{\pageref{firstpage}--\pageref{lastpage}} \pubyear{2002}

\maketitle

\label{firstpage}

\begin{abstract}
Forward shocks caused by the interaction between a relativistic \bw\ and the circum-burst medium 
are thought to be responsible for the afterglow emission in Gamma--Ray Bursts (GRBs). 
We consider the hydrodynamics of a spherical relativistic \bw\ expanding 
into the surrounding medium and we generalize the standard theory in order to account 
for several effects that are generally ignored.
In particular, we consider the role of adiabatic and radiative losses on the hydrodynamical 
evolution of the shock, under the assumption that the cooling losses are fast.
Our model can describe adiabatic, fully radiative and semi-radiative blast waves, 
and can describe the effects of a time-varying radiative efficiency. 
The equations we present are valid for arbitrary density profiles, 
and also for a circum--burst medium enriched with  electron--positron pairs. 
The presence of pairs enhances the fraction of shock energy gained by the leptons, 
thus increasing the importance of radiative losses.
Our model allows to study whether the high--energy ($>0.1\,$GeV) emission in GRBs 
may originate from afterglow radiation.
In particular, it is suitable to test whether the fast decay of the high--energy light curve 
observed in several Fermi LAT GRBs can be ascribed to an initial radiative phase, 
followed by the standard adiabatic evolution. 
\end{abstract}

\begin{keywords}
gamma-ray burst: general -- X-rays: general
\end{keywords}

\section{Introduction}

Afterglow radiation in Gamma--Ray Bursts (GRBs) is attributed to external shocks produced by the interaction between the ultra-relativistic ejecta and the 
circum--burst medium (CBM). 
As the collision evolves, a relativistic forward shock propagates into the 
surrounding medium, sweeping up and accelerating the ambient matter. 
Synchrotron emission from the shocked electrons powers the observed afterglow, 
which is usually detected in the X--ray, optical and sometimes radio bands.

Recent observations by the {\it Fermi}/LAT (Large Area Telescope; \citealt{atwood09}) 
and {\it AGILE}/GRID (Gamma--Ray Imaging Detector; \citealt{tavani09,barbiellini02}) 
telescope have revealed that a few percent of GRBs, both long and short, 
are detected in the MeV--GeV energy range, confirming previous findings by EGRET 
(the Energetic Gamma Ray Experiment Telescope onboard the Compton Gamma--Ray Observatory). 
By inspecting the temporal and spectral behavior above $0.1\,$GeV, 
some generic properties of the high-energy emission have been inferred 
\citep{ghisellini10,omodei09}: 
(\textit{i}) the high--energy radiation typically starts at early times, during the prompt phase, 
with a delay up to a few seconds after the beginning of the sub--MeV prompt emission;
(\textit{ii}) it lasts longer than the prompt keV--MeV emission detected by 
the {\it Fermi}/GBM;
(\textit{iii}) the high-energy flux decays as a power law in time, typically steeper than $t^{-1}$.
 
The emission mechanism remains controversial. 
A promising possibility, investigated by several authors, 
is that the high--energy emission has the same 
origin as the afterglow, at least when the MeV--GeV emission appears as a spectral component separate from the prompt, and it persists after the prompt has faded out \citep[][but see \citealt{piran10,maxham11} for caveats]{kumar09,kumar10,ghisellini10,ghirlanda10,corsi10,gao09}. 
The brightest bursts detected by LAT show a peak in their $>0.1\,$GeV light curve at early times. 
If interpreted as the onset of the afterglow emission, the peak time can be used to 
estimate the initial bulk \lf\ \go\ of the blast wave. 
The typical values obtained for the LAT bursts are quite large, around \go$\sim$10$^3$
\citep{ghisellini10,ghirlanda10}.
Beyond the peak, the $>0.1\,$GeV light curves decay with a temporal slope steeper 
than $t^{-1}$, which can be easily explained if at these times radiative losses 
affect the dynamical evolution of the blast wave, as suggested by \cite{ghisellini10}. 

Models where the energy lost to radiation is a sizable fraction of the blast wave energy are referred to as radiative. 
Radiative evolution sets in if the following two conditions are both satisfied: 
(\textit{i}) a considerable fraction of the  energy dissipated in external shocks is given to leptons 
(\epse$\simeq\,$$1$) {\it and} 
(\textit{ii}) leptons are in the fast cooling regime, i.e. \epsrad\  is close to unity. Here, \epsrad\ is defined as the fraction of the energy gained by electrons which is promptly radiated.
Estimates derived from modeling the X-ray and optical afterglow suggest that \epse\ may 
be roughly distributed around 0.1 \citep{panaitescu01}, which has been 
often adopted as a typical value. 
Even if our knowledge of particle acceleration in relativistic shocks is poor, 
it is reasonable to imagine that higher values of \epse\ could be achieved if the 
pre-shock medium is enriched with electron-positron pairs.
\citet{ghisellini10} considered the presence of pairs in the circum--burst 
medium to explain why the fireball may be radiative at early times, 
when the GeV emission takes place. 
A process that produces pairs in the ambient medium 
(invoking prompt photons scattered by the circum--burst electrons)
has been considered by
\cite{thompson00} and \cite{beloborodov02}.
This process is efficient up to some radius $R_{\rm load}$, 
beyond which the probability for an ambient electron to intercept a prompt 
photon becomes smaller than unity.

Even when the fireball is highly radiative in its early evolution, 
a transition to the standard adiabatic regime should occur when: 
(\textit{i}) the radiative process switches from fast to slow cooling 
(i.e. \epsrad\ decreases from $\sim1$ to \epsrad$\ll 1$); or 
(\textit{ii}) the physical conditions motivating large \epse\ cease to hold. 
For example, in the pair enrichment model described above, 
at radii larger than $R_{\rm load}$ the pair loading gets inefficient,
the pair--to--proton ratio drops and \epse\ decreases, 
driving the transition from the radiative to the adiabatic regime. 

At present, nearly 30 GRBs have been detected by LAT and the relation between the high--energy 
emission and the afterglow radiation can  
now be  investigated in more detail.
To probe the presence of an initially radiative epoch and to follow the transition 
 to the adiabatic regime as the pairs--to--proton ratio in the CBM decreases, we need to generalize the standard model for the hydrodynamical evolution 
of GRB external shocks.

\citet{blandford76} found a self--similar solution that describes the deceleration stage of 
adiabatic blast waves (i.e., in the case of negligible radiative losses), 
valid as long as the velocities are relativistic. 
Simpler analytical models based on the thin homogeneous shell approximation 
(where the hydrodynamical properties of the fluid behind the shock  are 
taken to be uniform inside the shell) allow to describe
the whole evolution of the shell \lf, from the coasting to the \nr\ epoch, but they suffer from some limitations. 
During the relativistic deceleration stage of adiabatic blast waves, they differ from the Blandford--McKee solution by a numerical factor \citep{panaitescu00}. Also, some important ingredients are often missing, 
such as the role of adiabatic expansion losses 
\citep{dermer01} or a proper description of the \nr\ phase \citep{huang99}. 
In all these approaches, the external medium density is usually considered to have a power--law radial profile and negligible pair loading.
The impact of a pair loaded medium on the observed afterglow radiation is exhaustively addressed by \citet{beloborodov05}.
In that work the author developed a Lagrangian formalism for afterglow calculations and showed that the radiation from the shocked pairs is likely to dominate the emission in the IR, optical, and UV bands during the first few minutes.

In this paper we develop a theoretical model able to describe:
\begin{itemize}
\item the whole blast wave evolution, starting from the initial coasting phase
($\Gamma=\Gamma_0=\rm const$) down to non--relativistic speeds;
\item adiabatic, fully radiative and semi--radiative regimes;
\item the transition from the radiative to the adiabatic regime, due to the 
change from fast to slow cooling or to the decrease in $\epsilon_e$;
\item adiabatic expansion losses;
\item a CBM characterized by an arbitrary density profile, possibly enriched with electron-positron pairs 
and pre--accelerated by the prompt radiation;
\item 
the evolution of the reverse shock.
\end{itemize}

Our model is based on the simple homogeneous shell approximation 
(see \citealt{piran99} for an exhaustive treatment of this approach). 
However, we obtain a more realistic description 
(\textit{i}) by recovering the Blandford \& McKee (1976)
self--similar solution 
during the relativistic deceleration phase of adiabatic blast waves; 
and (\textit{ii}) by properly accounting for the adiabatic losses. 
By including the adiabatic losses in a self-consistent way, we can capture the re--acceleration of the blast wave when the CBM density drops faster than $R^{-3}$, and the internal energy accumulated at earlier times is converted back into bulk kinetic energy. Also, we correctly recover the Sedov--Taylor solution \citep{sedov46,taylor50,vonneumann47} 
during the non--relativistic regime.

Our model allows to investigate more general issues, not necessarily 
related to the GeV emission. 
In particular, in this work we use our model to study: 
\begin{itemize}
\item the importance of radiative losses for arbitrary values of \epse;
\item the impact of an initial radiative phase on the energy content of the blast wave, and 
its effect on the subsequent adiabatic epoch  (and then on the X--ray and optical emission);
\item the relation between the initial bulk \lf\ \go\ and the peak 
time of the afterglow light curve.
\end{itemize}
In a follow--up paper (Nava et al., in preparation), we will test the external shock origin of the MeV--GeV emission, by simultaneously fitting the GeV, X--ray and optical light curves 
and spectra of the GRBs detected by LAT. 
We will adopt the dynamical model developed 
in this work and we will consider both the synchrotron 
and the synchrotron self--Compton processes, 
in order to predict light curves and spectra at different frequencies.
This study will probe whether an initial radiative 
phase is needed to explain the steep GeV light curves, and if the pair enrichment 
model described above can be a viable explanation for this early radiative epoch.

The presence of such a radiative phase could also affect the time when the blast wave becomes non--relativistic (which should occur earlier if the fireball efficiently 
loses energy in its early evolution). 
A detailed description of the transition to the \nr\ phase, accounting for 
the effects of lateral spreading in the realistic case of a collimated 
outflow, will be the subject of a follow--up paper (Sironi et al., in preparation).

In the following, primed quantities are measured in the frame comoving with the
fireball, unprimed quantities are measured in the frame of the progenitor star.
This will be true for all quantities except the particle momentum $p$ and the particles random
Lorentz factor $\gamma$, that are comoving
quantities, but that we leave unprimed for the ease of notation.

\section{Afterglow emission: the standard model}\label{stateoftheart}
It is possible to identify different stages in the evolution of a spherical relativistic blast wave.
After the acceleration phase,
the ejecta freely expand with constant \lf\ \go\ (coasting phase). 
When the internal energy of the shocked swept--up matter approaches the explosion 
energy (i.e., $\Gamma_0^2mc^2$$\,\sim\,$$E_0$), the blast wave enters a second stage, 
where both the shock wave and the shocked material decelerate, and the kinetic 
energy is gradually transformed into random internal energy and partially radiated. 
In the third stage of the evolution, the blast wave decelerates down to \nr\ velocities.
Most of the detected \ag\ radiation is emitted during the relativistic deceleration 
stage, via synchrotron cooling of the shocked electrons.

In this section, we briefly revise the methods that are commonly used 
to describe the evolution of external shocks in GRBs, and we emphasize 
the limitations of the different  approaches. 
In \S2.1 we review the self--similar solutions presented by \citet{blandford76} 
for adiabatic blast waves and by \citet{cohen98} for semi--radiative shells. 
These models describe the complete radial profile of the 
hydrodynamical quantities inside the shell.
In this context, \cite{beloborodov06} developed a mechanical model for relativistic blast waves and included the effects of the reverse shock.
Specifically, they made use of the jump conditions across both the reverse and the forward shock and applied the conservation of the mass, energy and momentum fluxes across the shocks.
The reverse shock vanishes as it crosses the end of the ejecta, and the blast wave then enters the self-similar stage where the forward
shock dynamics is described by the solution of \citet{blandford76}.

The validity of the \citet{blandford76} solution is limited to the 
relativistic deceleration phase. 
In contrast, in the so--called homogeneous shell approximation, the whole 
blast wave evolution can be followed, from the coasting to the \nr\ epoch, by considering a series of inelastic collisions 
between the shell and the external medium and by solving the equations for the 
energy and momentum conservation of the system. However,  
 the hydrodynamical properties of the fluid behind the shock are taken 
to be uniform inside the shell. 
In \S2.2 and 2.3 we present two 
common approaches based on the homogeneous shell approximation.

\subsection{Blandford \& McKee 1976 (BM76)}

\citet[][hereafter BM76]{blandford76} presented a fluid dynamical treatment 
of a relativistic spherical blast wave enclosed by a strong shock, 
under the assumption of adiabatic evolution. 
They considered the relativistic jump conditions across the shock and 
solved the continuity equations to derive the radial profile of the 
hydrodynamic quantities inside the shell. 
They derived a self--similar solution describing the deceleration stage of an 
adiabatic relativistic shock wave propagating into a medium with a power--law 
density profile $\rho(R)=A_0R^{-s}$, where $R$ is the distance of the shock 
from the center of the explosion. 
Their result represents the ultra--relativistic version of the Sedov--Taylor solution,
which describes the \nr\ stage of adiabatic shocks.

By definition, in the adiabatic regime the blast wave energy is constant and equal 
to the initial value $E_0$. 
By integrating the energy density of the shocked fluid over the shell 
volume, BM76 found:
\begin{equation}
E_0=\frac{12-4s}{17-4s}\Gamma^2 m c^2  = \frac{16\pi A_0 c^2}{17-4s} R^{3-s}\Gamma^2~,
\label{eq:bm}
\end{equation}
where $m$ is the ambient mass swept up at radius $R$, and $\Gamma$ is 
the bulk Lorentz factor of the fluid just behind the shock. 
This relation prescribes the dependence of $\Gamma$ on $R$, giving $\Gamma\propto R^{-(3-s)/2}$. 
In the two common cases (homogeneous density, $s=0$; and wind-like profile, $s=2$), 
it gives  $\Gamma\propto R^{-3/2}$ and $\Gamma\propto R^{-1/2}$, respectively. 
Even if this approach provides a reliable description of the deceleration epoch 
for adiabatic blast waves, both the coasting and the non--relativistic phases 
(and the transition between different stages of the evolution) cannot be properly described.

BM76 also investigated the case of fully radiative blast waves. 
In this regime, they considered all the shocked matter to be concentrated 
in a cold, thin and homogeneous shell. 
From energy and momentum conservation, they found that the evolution of 
the shell Lorentz factor $\Gamma$ satisfies
\begin{equation}
\Gamma-1=2\left[\frac{(m+M_0)^2(\Gamma_0+1)}{M_0^2(\Gamma_0-1)}-1  \right]^{-1}~,
\label{eq:bmrad}
\end{equation}
where $M_0$ is the initial mass of the ejecta.
Contrary to the BM76 adiabatic solution in \eq{bm}, this equation 
(based on the homogeneous shell approximation) is 
also valid in the coasting epoch and for \nr\ velocities. 
It is the ultra--relativistic generalization of the momentum--conserving 
snowplow \citep{ostriker_mckee88}, characterized by a thin shell 
where all of the matter is concentrated, with an empty cold interior. 
As pointed out by \citet{cohen98}, the solution in \eq{bmrad} has been derived without 
accounting for the factor $4/3$ that comes from the Lorentz transformation of the energy density in a relativistically hot fluid.

\citet{cohen98} have looked for self--similar solutions to describe the relativistic 
deceleration stage of semi--radiative blast waves, i.e. the generalization of the BM76 
solution (\eq{bm}), which only holds  for adiabatic shells.
They consider an adiabatic shock followed by a narrow radiative region and a self--similar 
adiabatic interior.  
In their approach, the shell cavity is hot and pressurized, as opposite to the empty 
and cold interior of the radiative solution by BM76 in Eq. \ref{eq:bmrad}. 
In the fully radiative regime, their solution does not reduce to 
the momentum--conserving snowplow of BM76, but it rather represents the 
ultra--relativistic generalization of the pressure-driven snowplow solution \citep{ostriker_mckee88}.

\subsection{Piran 1999 (P99)}
\citet[][hereafter P99]{piran99} solved for the evolution of spherical blast waves 
in the homogeneous shell approximation, with a generic radiative efficiency.
The evolution of the \bw\ Lorentz factor is derived by imposing energy and momentum 
conservation in the collision between the blast wave and the external matter. 
An analytic equation for $\Gamma(R)$ (from the coasting to the 
non--relativistic phase) can be derived.  
This method suffers from some limitations.
In the adiabatic regime, the predicted scaling law between $\Gamma$ and $R$ 
during the relativistic deceleration stage is the same as derived by BM76, 
but the two solutions differ by a numerical factor \citep{panaitescu00}.
Moreover, adiabatic expansion losses are not considered. 
For non--relativistic velocities, Piran's solution gives $\beta\propto R^{-3}$. 
As \citet{huang99} pointed out, this scaling does not agree with the 
Sedov--Taylor solution ($\beta\propto R^{-3/2}$), which is expected to describe 
the final \nr\ stages of the fireball evolution.
In the fully radiative case, P99 recovers the same solution derived by 
BM76 (Eq. \ref{eq:bmrad}), which describes a momentum--conserving snowplow.

\subsection{Huang et al. 1999 (H99)}
In order to derive a relation between $\Gamma$ and $R$ valid also in the \nr\ regime, 
\citet[][hereafter H99]{huang99} considered the homogeneous shell model of P99 
but imposed a different equation to describe the internal energy of the fireball. 
Their assumption states that the comoving internal energy of adiabatic 
blast waves should be $E'_{int}=(\Gamma-1)\,m\,c^2$.\footnote{For fully radiative blast 
waves, $E'_{int}=0$, and their solution coincides with the momentum--conserving 
snowplow of BM76 and P99.}

This prescription for the internal energy mimics the role of adiabatic losses. The Rankine--Hugoniot jump conditions state that the average kinetic energy per unit mass is constant across a relativistic shock  (as measured in the post-shock frame), meaning that the random Lorentz factor of an element $dm$ of shocked material equals the shell Lorentz factor $\Gamma(m)$ at that time. At early times, when the shell Lorentz factor was larger, the material was shocked to hotter temperatures. It follows that, in the absence of adiabatic and radiative losses, the internal energy of the shell should be $E'_{int}=\int[\Gamma(m)-1]\,dm\,c^2$. By imposing $E'_{int}=(\Gamma-1)\,m\,c^2$, H99 implicitly mean that the effect of adiabatic losses is to keep the random Lorentz factor of all the swept--up matter equal to the current bulk \lf\ of the shell, regardless of the hotter temperatures of the material accreted at earlier times. Although this treatment of the adiabatic losses relies on an ad-hoc prescription, it allows to recover the correct scaling between $\beta$ and $R$ predicted by the Sedov-Taylor solution, $\beta\propto R^{-3/2}$.

A more accurate treatment of the adiabatic losses, giving the correct scaling 
of $\beta(R)$ in the \nr\ regime without any ad-hoc assumption, has been considered 
by \citet{panaitescu98} and \citet{dermer01}, but only for  
adiabatic shocks in power--law density profiles. 
In our model, that we describe below, we generalize their formalism to the case of 
semi--radiative blast waves propagating in a medium with an arbitrary density profile.

\section{Our model}
\label{ourmodel}

In this section, we propose a novel method to describe the evolution of spherical 
relativistic  blast waves, in the homogeneous shell approximation. 
Our model overcomes the problems and limitations present in other approaches, 
that we have summarized in \S\ref{stateoftheart}. 
In particular, our method is able to describe: 
\begin{itemize} 

\item the coasting, deceleration and \nr\ phases, and the transition between these regimes. 
For adiabatic blast waves, the relativistic deceleration phase reproduces the BM76 solution, 
and in the \nr\ regime the shell velocity scales as $\beta\propto R^{-3/2}$, as predicted by 
the Sedov--Taylor solution.

\item  adiabatic, semi--radiative and fully radiative blast waves.

\item the case of a time--varying radiative efficiency, either resulting from a change 
in the shock microphysics (i.e., in the fraction \epse\ of shock energy transferred to 
the emitting leptons, or due to the transition from fast to slow cooling, which 
determines a decrease in the electron radiative efficiency \epsrad).

\item arbitrary density profiles (e.g., resulting from 
the interaction of the progenitor stellar wind with the surrounding interstellar medium), 
including the case of a circum--burst medium enriched with electron--positron pairs.

\item possible re--acceleration phases, caused by the conversion of part of the internal 
energy back into bulk kinetic energy. Re--acceleration is expected to occur if the blast wave encounters a sudden drop in the density of the external matter (i.e., the density profile becomes steeper than $\rho\propto R^{-3}$).

\end{itemize}
The model presented in this section does not account for the reverse shock nor for the pre--acceleration of the external medium. However, both processes might produce relevant effects on the early light curve of the afterglow emission \citep{beloborodov05,beloborodov06}. For this reason, in Appendix \ref{reverse shock} we modify the equations presented in this section in order to include the effects of the reverse shock, while, in Appendix \ref{preacceleration}, we show how it is possible to account for an external medium which has been pre--accelerated by the prompt radiation.

We now derive the equations describing the evolution of the shell bulk Lorentz factor 
$\Gamma$ as a function of the shock radius $R$. 

\vskip 0.2 cm
\noindent
{\it Total energy ---}
The energy density in the progenitor frame can be obtained from the Lorentz 
transformation: $e=(e'+P')\Gamma^2-P'$, where the pressure $P'$ is 
connected to the comoving energy density $e'$ and the comoving mass density $\rho'$ 
by the equation of state  $P'=(\adiab-1)(e'-\rho'c^2)$. 
Here, $\hat{\gamma}$ is the adiabatic index of the shocked plasma, which we parameterize 
as $\hat{\gamma}=(4+\Gamma^{-1})/3$ for the sake of simplicity, obtaining the 
expected limits $\hat{\gamma}\simeq4/3$ for $\Gamma\gg1$ and $\hat{\gamma}\simeq5/3$ for 
$\Gamma\rightarrow1$ \cite[see][for a more accurate prescription]{peer12}. 
The total energy in the progenitor frame will be $E=e V=eV'/\Gamma$, where $V$ is 
the shell volume in the progenitor frame, and $V'=\Gamma V$ in the comoving frame. 
The total energy of the shell in the progenitor frame is the sum of the 
bulk kinetic energy and the internal energy: 
\be
\label{eq:Etot}
E_{tot}=\Gamma(M_0+m)\,c^2+\Gamma_{\it eff} E'_{int}~,
\ee
where we have defined 
\be
\label{eq:gammaeff}
\Gamma_{\it \it eff}\equiv\frac{\adiab\Gamma^2-\adiab+1}{\Gamma}~,
\ee
to properly describe the Lorentz transformation of the internal energy. 
Here, $M_0+m=\rho' V'$ is the sum of the ejecta mass $M_0=E_0/\Gamma_0c^2$ and of the swept--up 
mass $m(R)$, and $E'_{int}=(e'-\rho'c^2) V'$ is the comoving internal energy. 
As pointed out by \citet{peer12}, the majority of current models use 
$\Gamma$ instead of $\Gamma_{\it eff}$, with an error up to a factor 
of $4/3$ in the ultra--relativistic limit, when $\Gamma_{\it eff}\rightarrow (4/3) \Gamma$.
Note that when accounting for the reverse shock the equation expressing the total energy of the shell
must be modified as described in Appendix \ref{reverse shock}. In particular, at early times only a fraction 
of the ejecta mass has been 
shocked and travels with bulk Lorentz factor $\Gamma$, while the remaining fraction moves
with $\Gamma_0$. When the reverse shock has crossed the ejecta, all the ejecta mass has been 
decelerated and Eq. \ref{eq:Etot} can be applied. In addition, the reverse shock will 
heat the ejecta, and the internal energy of the shocked ejecta must be considered. 
The changes to our dynamical model in the presence of a reverse shock are given 
in Appendix \ref{reverse shock}.

\vskip 0.2 cm
\noindent
{\it Formal equation for $\Gamma(R)$ ---}
The blast wave energy $E_{tot}$ in Eq. \ref{eq:Etot} can change due to ({\it i}) the rest 
mass energy $dm\,c^2$ accreted from the CBM, and ({\it ii}) the  energy lost to 
radiation $dE_{rad}=\Gamma_{\it eff}\,dE'_{rad}$.
If the external medium has been pre--accelerated by the interaction with the prompt radiation, 
the mass swept up by the forward shock is not at rest in the progenitor frame. 
In this section we neglect the possible motion of the CBM 
in the progenitor frame and defer to Appendix \ref{preacceleration} the description 
of how this effect can be easily included in the model.
The equation of energy conservation in the progenitor frame then reads
\be
d\left[\Gamma(M_0+m)c^2+\Gamma_{\it eff} E'_{int}\right]= dm\,c^2+\Gamma_{\it eff} dE'_{rad}~.
\label{eq:energy conservation}
\ee
The overall change in the comoving internal energy $dE'_{int}$ results from the sum of 
three contributions:
\be
dE'_{int} = dE'_{sh}+dE'_{ad}+dE'_{rad}~~.
\ee
The first contribution, $dE'_{sh}=(\Gamma-1) \,dm\,c^2$, is the random kinetic energy 
produced at the shock as a result of the inelastic collision with an element $dm$ of 
circum--burst material\footnote{As pointed out by BM76, in the post--shock frame,
the average kinetic energy per unit mass is constant across the shock, and equal to 
$(\Gamma-1) \,c^2$.}. 
The second term, $dE'_{ad}$, is the energy lost due to adiabatic expansion.
The adiabatic losses lead to a conversion of random energy back to bulk kinetic energy.
This contribution is generally negligible, but it may become important when the CBM density decreases faster than $\rho\propto R^{-3}$, as shown in \S\ref{reacceleration}.
The third term, $dE'_{rad}$, accounts for radiative losses. 

From Eq. \ref{eq:energy conservation}, we derive the equation for the evolution of 
the shell Lorentz factor:
\be
\label{eq:G solve}
\frac{d\Gamma}{dR}=-\frac{(\Gamma_{\it eff}+1)(\Gamma-1)\,c^2\frac{dm}{dR}
+\Gamma_{\it eff}\frac{dE'_{ad}}{dR}}{(M_0+m)\,c^2+E'_{int}\frac{d\Gamma_{\it eff}}{d\Gamma}}~,
\ee
which reduces to Eq. 5 of \citet{dermer01} if we replace $\Gamma_{\it eff}$ with $\Gamma$. 
For a spherical shell, $dm/dR=4\pi R^2 \rho$, where $\rho(R)$ is the circum--burst mass density. The term $\Gamma_{eff} dE'_{ad}/dR$, accounting for adiabatic losses, allows to
describe the re--acceleration of the fireball when the density
of the CBM is a fast decreasing function of $R$
(see \S\ref{reacceleration} and the Appendix A). 

\vskip 0.2 cm
\noindent
{\it Specifying $dE'_{ad}$ and $E'_{int}$ ---}
Eq. \ref{eq:G solve} is the formal equation giving $\Gamma(R)$.
To solve it, we need to specify $dE'_{ad}$ and $E'_{int}$. 
To this aim, we adopt the approach proposed by \citet{dermer01} 
and we generalize their method to account for 
({\it i}) arbitrary profiles of the circum--burst density $\rho(R)$, and ({\it ii}) a  
time--varying overall radiative efficiency $\epsilon(R)$. 
The efficiency $\epsilon$ is 
defined as follows: a fraction \epse\ of the energy dissipated by the shock $dE'_{sh}$ 
is gained by the leptons, which then radiate a fraction  \epsrad\ of their internal energy.
It follows that the energy lost to radiation is  $dE'_{rad}=-\epsilon_{rad}\epsilon_{e}\, dE'_{sh}=-\epsilon \,dE'_{sh}$, with 
$\epsilon\equiv\,$\epsrad\epse. 
In other words, $\epsilon$ is the overall fraction of the shock--dissipated energy that goes into radiation.

After gaining a fraction \epse\ of the shock energy, the mean random Lorentz factor of 
post--shock leptons becomes $\gamma_{acc,e}-1= (\Gamma-1)\epsilon_{e}/\mu_e$. 
Here, $\mu_{e}=\rho_e/\rho$ is the ratio between the mass density $\rho_e$ of shocked electrons and positrons 
(simply ``electrons'' from now on) and the total mass density of the shocked matter $\rho$.  
In the absence of electron--positron pairs, electrons and protons have the same number 
density and the same radial profile, and $\mu_e$ is simply the ratio between 
the electron and proton mass: $\mu_e=m_{e}/(m_e+m_{p})\simeq m_{e}/m_{p}$. 
We assume that, right behind the shock, the freshly shocked electrons instantaneously 
radiate a fraction \epsrad\ of their internal energy, so that their mean random \lf\ 
decreases down to 
\be\label{eq:gerad}
\gamma_{rad,e}-1=(1-\epsilon_{rad})(\gamma_{acc,e}-1)=
(1-\epsilon_{rad})(\Gamma-1)\frac{\epsilon_{e}}{\mu_e}~.
\ee

Following this transient, electrons cool only due to adiabatic losses. 
The assumption of instantaneous radiative losses is verified in the fast cooling regime 
($\epsilon_{rad}\sim1$), which is required (but not sufficient) to have $\epsilon\sim1$ 
(i.e., a fully radiative blast wave). 
In the opposite case $\epsilon_{rad}\ll1$, the evolution is nearly adiabatic ($\epsilon\ll1$), 
regardless of the value of \epse, 
and the details of the radiative cooling processes are likely to be unimportant for the shell dynamics. 
The case with intermediate values of \epsrad\ and $\epsilon$ is harder to treat analytically, 
since the electrons shocked at radius $R$ may continue to emit copiously also at larger distances, 
affecting the blast wave dynamics. 
In the following, we implicitly assume either that  the emitting electrons are in the fast cooling 
regime ($\epsilon_{rad}\sim1$) or that the radiative losses are unimportant for the shell dynamics (i.e., $\epsilon\ll1$).

Assuming that protons gain a fraction \epsp\ of the energy dissipated by the shock 
(with $\epsilon_p=1-\epsilon_e-\epsilon_B\simeq1-\epsilon_e$, if the fraction 
$\epsilon_B$ converted into magnetic fields is negligible), 
their mean post--shock Lorentz factor will be
\be\label{eq:gpacc}
\gamma_{acc,p}-1=(\Gamma-1)\frac{\epsilon_p}{\mu_p}~,
\ee
where $\mu_p=\rho_p/\rho$ is the ratio between the mass density of shocked protons $\rho_p$ and the total shocked mass density $\rho$. 
In the standard case, when pairs are absent, $\mu_p\simeq1$. 
Since the proton radiative losses are negligible, the shocked protons will lose their  
energy only due to adiabatic cooling. 

\vskip 0.2 cm
\noindent
{\it Comoving volume and adiabatic losses ---}
To estimate the adiabatic losses, we assume that the shell comoving volume scales 
as $V'\propto R^3/\Gamma$, corresponding to a shell thickness in the progenitor frame $\sim R/\Gamma^2$. 
This scaling is correct for both relativistic and non--relativistic shocks, in the decelerating phase (BM76). 
For re--accelerating relativistic shocks, \citet{shapiro80} showed that the thickness of the 
region containing most of the blast wave energy is still $\sim R/\Gamma^2$. 
For the sake of simplicity, we neglect changes in the comoving volume due to a time--varying 
adiabatic index or radiative efficiency. 

The radial change of the comoving momentum $p$ of a shocked particle, as a result of expansion 
losses, will be governed by
\be
\label{eq:p_ad}
\left(\frac{dp}{dr}\right)_{ad}=-\frac{p}{3}\frac{d\ln V'}{dr}
=-\frac{p}{3}\left(\frac{3}{r}-\frac{d\ln \Gamma}{dr}\right)~,
\ee
which is valid both in the relativistic and non--relativistic limits \citep{dermer01}. 
The comoving momentum at radius $R$, for a particle injected with momentum $p(r)$ when 
the shock radius was $r$, will be
\be
p_{ad}(R,r)=\frac{r}{R}\left[\frac{\Gamma(R)}{\Gamma(r)}\right]^{1/3}p(r)~.
\label{pad}
\ee
The comoving momentum $p(r)$ is given by  $p=({\gamma}^{2}-1)^{1/2}$, where for electrons 
$\gamma=\gamma_{rad,e}(r)$ (\eq{gerad}), and for protons $\gamma=\gamma_{acc,p}(r)$ (Eq. \ref{eq:gpacc}). 
We remind that, by assuming that the shocked electrons promptly radiate at the shock, 
and then they evolve adiabatically, we are implicitly considering only the fast cooling 
regime $\epsilon_{rad}\sim1$. Of course, our treatment is also valid in the quasi-adiabatic regime $\epsilon\ll1$, when the radiative losses do not affect the shell dynamics.

\vskip 0.2 cm
\noindent
{\it Internal energy and adiabatic losses ---}
Considering the proton and lepton energy densities separately,
the comoving internal energy at radius $R$ will be
\be
\label{eq:int energy}
E'_{int}(R)\!=\!4 \pi c^2\!\!\! \int_0^R\!\!\! 
dr r^2 \left\{\rho_{p}(r) [\gamma_{ad,p}(R,r)\!-\!1]\!+\!\rho_{e}(r) 
[\gamma_{ad,e}(R,r)\!-\!1] \right\}\!\!\!
\ee
where $\gamma_{ad}=({p}_{ad}^{2}+1)^{1/2}$.
With the help of Eq. \ref{pad}, we can explicitly find $E'_{int}(R)$ and insert it
in Eq. \ref{eq:G solve}.

\vskip 0.2 cm
\noindent
The other term needed in Eq. \ref{eq:G solve} is $dE'_{ad}/dR$.
First, we derive $(d\gamma/dR)_{ad}$ for a single particle, using $\gamma=(p^{2}+1)^{1/2}$ and \eq{p_ad}. Then we integrate over the total number of particles, again considering 
protons and leptons separately.
Finally we have:
\be
\label{eq:ad losses}
\frac{dE'_{ad}(R)}{dR}&=&-4 \pi c^2 \left(\frac{1}{R}-\frac{1}{3}
\frac{d \log \Gamma}{dR}\right)\int_0^R dr r^{2} 
\times \nonumber \\  
&& \left\{\rho_{p}(r) \frac{{p}^2_{ad,p}(R,r)}{{\gamma_{ad,p}}(R,r)}+\rho_{e}(r)  
\frac{{p}^2_{ad,e}(R,r)}{{\gamma_{ad,e}}(R,r)}\right\}~.
\ee
In Eqs. \ref{eq:int energy} and \ref{eq:ad losses}, we have assumed that only the swept--up 
matter is subject to adiabatic cooling, i.e., that the ejecta particles are cold. 

\vskip 0.2 true cm
\noindent
{\it Relativistic phase ---}
As long as the shocked particles remain relativistic (i.e., $\gamma_{ad}\simeq p_{ad}\gg1$), 
the equations for the comoving internal energy and for the adiabatic expansion losses assume 
simpler forms:
\be
 \label{eq:eint_rel}
E'_{int}(R)&\!\!\!\!=\!\!\!\!&4 \pi c^2\!\!\! \int_0^R\!\! 
dr r^2 \frac{r}{R}\left[\frac{\Gamma(R)}{\Gamma(r)}\right]^{1/3} 
\Gamma(r)\left\{ \frac{\epsilon_p}{\mu_p}\rho_{p}+
(1-\epsilon_{rad})\frac{\epsilon_e}{\mu_e}\rho_{e}\right\} \nonumber\\
&\!\!\!\!=\!\!\!\!&4 \pi c^2\!\!\!\int_0^R\!\! dr r^2 \frac{r}{R}
\left[\frac{\Gamma(R)}{\Gamma(r)}\right]^{1/3} \Gamma(r)\,\rho(r)\left(\epsilon_p+\epsilon_e-\epsilon \right)\\
\frac{dE'_{ad}(R)}{dR}&\!\!\!\!=\!\!\!\!&-
E'_{int}(R)\left(\frac{1}{R}-\frac{1}{3}\frac{d \log \Gamma}{dR}\right)~
\ee
In the second line of \eq{eint_rel} we have used that $\epsilon=\epsilon_e\epsilon_{rad}$. 
In the absence of significant magnetic field amplification, $\epsilon_p+\epsilon_e\simeq1$ so that $\epsilon_p+\epsilon_e-\epsilon\simeq1-\epsilon$, and the radiative processes of the blast wave are entirely captured by the single efficiency parameter $\epsilon$.
In the fast cooling regime $\epsilon_{rad}\sim1$ and $\epsilon\simeq\epsilon_e$. 
In this case the term 
$\epsilon_p+\epsilon_e-\epsilon$  reduces to \epsp, meaning that, regardless of the 
amount of energy gained by the electrons, in the fast cooling regime the adiabatic losses are dominated by the protons, since the electrons lose all their energy to radiation. 

\vskip 0.2 cm
\noindent
{\it Adiabatic regime ---}
Evaluating these expressions for adiabatic blast waves in a power--law density profile 
$\rho\propto R^{-s}$, we obtain
\be
E'_{int}(R)=\frac{9-3s}{9-2s}\Gamma m c^2~~;~~~~~~  
\frac{dE'_{ad}(R)}{dR}=-\frac{(9-s)(3-s)}{2(9-2s)}\frac{\Gamma m c^2}{R} 
\ee
where we have used that $\Gamma\propto R^{-(3-s)/2}$ as in the adiabatic BM76 solution. 
\vskip 0.2 cm
\noindent
{\it Fully radiative regime ---}
In the fully radiative regime $\epsilon=1$, which implies $E'_{int}=0$ and $dE'_{ad}=0$, 
Eq.~\ref{eq:G solve} reduces to
\be
\label{eq:rad}
\frac{d\Gamma}{dm}=-\frac{(\Gamma_{\it eff}+1)(\Gamma-1)}{M_0+m}~~,
\ee
which describes the evolution of a momentum--conserving (rather than pressure--driven) snowplow. 
If we replace $\Gamma_{\it eff}\rightarrow \Gamma$, the solution of this equation coincides with the result 
by BM76 (i.e., our Eq. \ref{eq:bmrad}) and also with the equations derived by P99 and H99, since for fully radiative solutions the specific treatment of the adiabatic losses is unimportant. 

\vskip 0.2 cm
\noindent
{\it Correction factor ---}
Since our model is based on the homogeneous shell approximation, our adiabatic solution does not 
recover the correct normalization of the BM76 solution (see Eq. \ref{eq:bm}). 
In our treatment, the total energy of a relativistic decelerating adiabatic blast 
wave in a power--law density profile $\rho(R)\propto R^{-s}$ is
\be
E_0\simeq \Gamma_{\it eff} E'_{int}\simeq\frac{4}{3}\Gamma E'_{int}\simeq\frac{12-4s}{9-2s}\Gamma^2 m c^2~~,
\ee
so that the BM76 normalization can be recovered if we multiply the density of external 
matter in Eqs. \ref{eq:G solve}, \ref{eq:int energy} and \ref{eq:ad losses} 
by the factor $(9-2s)/(17-4s)$.\footnote{This correction is only appropriate for 
a power--law density profile.} 
To smoothly interpolate between the adiabatic regime (where we want to recover the BM76 solution) and the radiative regime, 
we suggest a correction factor:
\be
\label{eq:corr}
C_{BM76,\epsilon}\equiv\epsilon+\frac{9-2s}{17-4s}(1-\epsilon)~~.
\ee
No analytic model exists that properly captures the transition between an adiabatic relativistic 
blast wave and the momentum--conserving snowplow, as $\epsilon$ increases from zero to unity 
(instead, the model by \citealt{cohen98} approaches the pressure--driven snowplow in 
the limit $\epsilon\rightarrow1$). 
For this reason, we decide to employ the simple interpolation in Eq.~\ref{eq:corr}, in order to join the fully adiabatic BM76 solution with the fully radiative momentum-conserving snowplow.

\vskip 0.2 cm
In summary, Eqs.~\ref{eq:G solve}, \ref{eq:int energy} and \ref{eq:ad losses}, 
complemented with the correction in Eq.~\ref{eq:corr} (which should by applied to every occurrence of  
external density and external matter) completely determine the evolution of the shell Lorentz factor $\Gamma$ 
as a function of the shock radius $R$.

In our model it is easy to recover the solutions by P99 and H99. 
First of all, we should replace $\Gamma_{\it eff}\rightarrow\Gamma$ and $C_{BM76,\epsilon}\rightarrow 1$. 
The approach described by P99 can be recovered for $dE'_{ad}=0$, so  
$dE'_{int}=(1-\epsilon)(\Gamma-1)dm\,c^2$, while 
the equation given by H99 can be obtained for $E'_{int}=(1-\epsilon)(\Gamma-1)mc^2 $, 
which implies $dE'_{ad}=(1-\epsilon)d\Gamma mc^2$, at odds with respect to our Eq.~\ref{eq:ad losses}.

\subsection{Testing the model}
Before applying our model to the physics of GRB afterglows, it is healthy to compare it with previous studies
and to understand the differences.

\begin{figure}
\hskip -0.1 truecm
\includegraphics[scale=0.43]{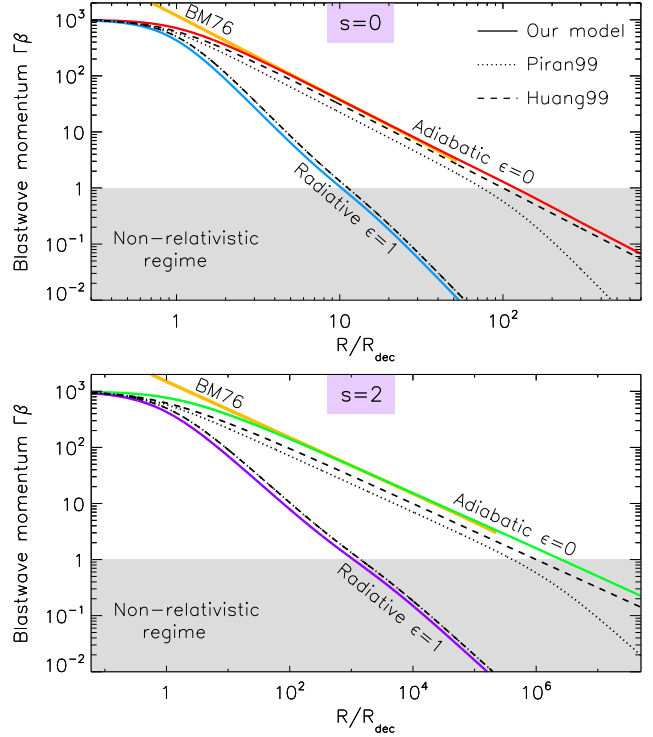}
\caption{
Evolution of the shell momentum $\Gamma \beta$ as a function of the shock radius $R$, 
normalized to the deceleration radius $R_{dec}$, in the case of homogeneous ISM density 
($s=0$, upper panel) and wind--like profile ($s=2$, bottom panel). 
Solid curves show the prediction of our model for the fully adiabatic and fully radiative solutions. 
Dotted (dashed) curves show instead the P99 (H99) solution. 
The straight (yellow) line is the prediction by BM76, which is valid after the deceleration 
started and till the velocities are relativistic. The grey area indicates the non-relativistic regime.
}
\label{fig:momentum}
\end{figure}
\vskip 0.2 cm
\noindent
{\it Momentum of the blast wave: adiabatic and fully radiative cases}
Fig. \ref{fig:momentum} shows the evolution of the shell momentum $\Gamma\beta$ 
for a CBM density profile $\rho=A_0\,R^{-s}$. 
We compare our model (solid curves) with the predictions by P99 (dotted) and H99 (dashed), 
for adiabatic and fully radiative blast waves, for homogeneous ($s=0$, top panel) or wind
($s=2$, bottom panel) density profiles.
The shock radius is normalized to the deceleration radius, which is defined such that 
the swept--up mass is $m_{dec}\equiv m(R_{dec})=M_0/\Gamma_0$, so
\be
R_{dec}\equiv\left[\frac{(3-s)M_0}{4\pi A_0\Gamma_0}\right]^{1/(3-s)}~~.
\label{eq:rdec}
\ee 
In the relativistic adiabatic regime, \fig{momentum} shows that our solution reproduces 
the adiabatic relativistic solution of BM76 (yellow straight lines) for both the uniform and the 
wind density profile.
P99 and H99 underestimate the shell momentum with respect to the BM76 solution. 
Moreover, the approach by P99 does not correctly reproduce the Sedov--Taylor solution 
(with $\beta\propto R^{-3/2}$), since it gives $\beta\propto R^{-3}$ 
(as anticipated in \S\ref{stateoftheart}). 
For fully radiative blast waves, the solutions by P99 and H99 coincide and differ 
from our solution only because they use $\Gamma$ instead of $\Gamma_{\it eff}$ in \eq{rad}. 

\begin{figure}
\hskip +0.2 truecm
\includegraphics[scale=0.53]{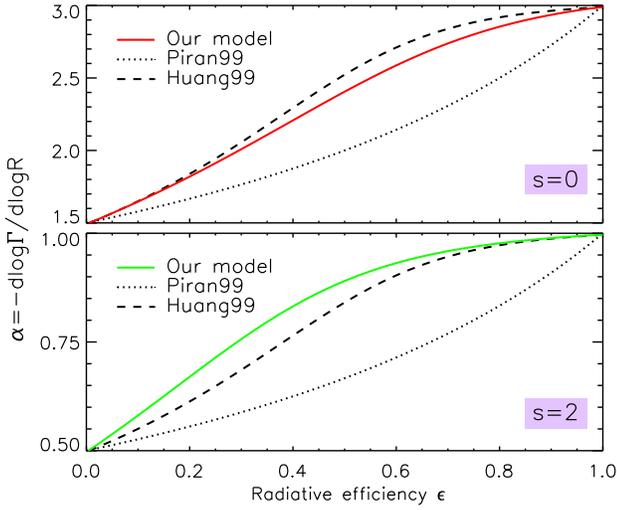}
\caption{
Comparison of the slope $\alpha$ (such that $\Gamma\propto R^{-\alpha}$) 
in our model (solid lines) with the predictions by P99 (dotted) and H99 (dashed), 
for different values of the radiative efficiency $\epsilon$. 
The slope is measured during the relativistic deceleration phase of evolution. 
The top panel refers to the case of a homogeneous interstellar medium, the bottom 
panel to a wind density profile. We remark that the slopes in our model do not depend on the correction factor $C_{BM76,\epsilon}$ in Eq.~\ref{eq:corr}.
}
\label{fig:slopes}
\end{figure}

As long as we are in the relativistic regime, \fig{momentum} shows that
all models agree on the slope of $\Gamma\beta$ as a function of $R$, both for the adiabatic and for the fully 
radiative case.

\vskip 0.2 cm
\noindent
{\it Momentum of the blast wave: semi--radiative case ---}
\fig{slopes} shows the slopes $\alpha\equiv-d\ln \Gamma/d\ln R$ as a 
function of the overall radiative efficiency $\epsilon$.
All three methods give $\alpha=(3-s)/2$ for $\epsilon=0$ and $\alpha=3-s$ for 
$\epsilon=1$, but they differ for intermediate values of $\epsilon$ (i.e., they predict different 
slopes in the case of semi--radiative blast waves). 
In this case the value of $\alpha$ in the model by P99 can be derived analytically, and it gives $\alpha=(3-s)/(2-\epsilon)$. 
In our model, an analytic expression for $\alpha$ can only be obtained for $\epsilon\ll1$, 
and it yields $\alpha=(9-3s+3\epsilon)/(6-4\epsilon)$. The same is true for the approach 
by H99, which results in $\alpha=(3-s)/(2-2\epsilon)$ for $\epsilon\ll1$. 
From \fig{slopes}, it is clear that the model by P99, which neglects the adiabatic 
expansion losses, always gives the flattest slopes
(i.e., the shell evolution remains close to the adiabatic case, everything else being fixed). 
Comparing our model with H99 we see that our model yields flatter slopes for a homogeneous medium 
(top panel), whereas the opposite holds for a wind profile (bottom panel).

\begin{figure}
\hskip -0.1 truecm
\includegraphics[scale=0.43]{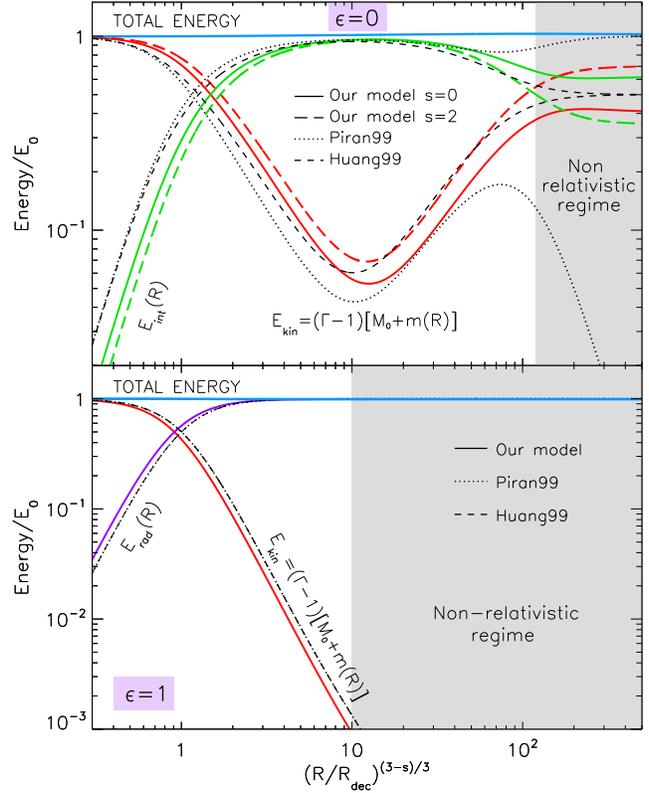}
\caption{
Evolution of the different contributions to the blast wave energy, normalized 
to the initial energy $E_0$. 
Solid (long dashed) lines refer to our model in the case of $s=0$ ($s=2$).
Dotted and dashed curves refer to P99 and H99, respectively. 
The top panel shows the adiabatic case $\epsilon=0$, the bottom panel presents 
the fully radiative case $\epsilon=1$. 
The contribution of the bulk kinetic  energy $E_{kin}=(\Gamma-1)(M_0+m)c^2$ 
is shown in red, the internal energy $E_{int}=\Gamma_{\it eff}E'_{int}$in green,  
the energy lost to radiation $E_{rad}=\int \Gamma_{\it eff} \,dE'_{rad}$ in purple. 
The sum of the different contributions is the cyan curve, demonstrating the  
conservation of energy. The grey area indicates the non-relativistic regime.
}
\label{fig:energy}
\end{figure}

\vskip 0.2 cm
\noindent
{\it Partition of energy ---}
\fig{energy} shows the partition of energy among different components 
(red for kinetic energy, green for internal energy, purple for radiated energy), 
during the evolution of the blast wave. 
The results of our model are plotted with solid and long--dashed lines, whereas 
the predictions by P99 and H99 are shown with dotted and dashed lines, respectively. 
In the coasting phase ($R\lesssim R_{dec}$), the blast wave energy is still dominated 
by the bulk kinetic energy of the ejecta (red curves). 
At $R\sim R_{dec}$, bulk and internal energies are comparable 
by definition.
For $R\gtrsim R_{dec}$, the ejecta decelerate, and the contribution of the bulk motion 
to the overall energy budget decreases. 

For adiabatic blast waves (top panel), the decrease in kinetic energy 
(red lines) is balanced by an increase in internal energy (green lines). 
After the shell has accreted a mass $m\sim M_0$, the blast wave kinetic energy 
becomes dominated by the swept--up material, rather than by the initial ejecta mass. 
This leads $E_{kin}$ to increase as 
$E_{kin}\sim \Gamma m c^2\propto m^{1/2}$. 
In the non--relativistic regime (shaded area), the internal energy 
$E_{int}=\Gamma(\Gamma-1)m\,c^2\sim(\Gamma-1)m\,c^2$ and the bulk kinetic energy 
$E_{kin}=(\Gamma-1)(M_0+m)c^2\sim(\Gamma-1)m c^2$ asymptote to the same value
in the model by H99 (dashed lines). 
In our model
the internal energy dominates over the kinetic energy for a uniform medium (solid lines), 
whereas the opposite holds for a wind--like density profile (long--dashed lines). 
Finally, we remark that in the model by P99 (dotted lines), that neglects the adiabatic 
expansion losses, most of the energy at late times is still in the form of internal energy, 
with a negligible contribution by the bulk kinetic energy.

For radiative blast waves (bottom panel), the kinetic energy (red line) is lost to 
radiation (purple line). 
The blast wave is cold (i.e., the contribution by the internal energy is negligible), since 
the random energy generated at the shock is promptly radiated.

\subsection{Complex density profiles and re--acceleration}
\label{reacceleration}
\begin{figure}
\hskip -0.4 truecm
\includegraphics[scale=0.44]{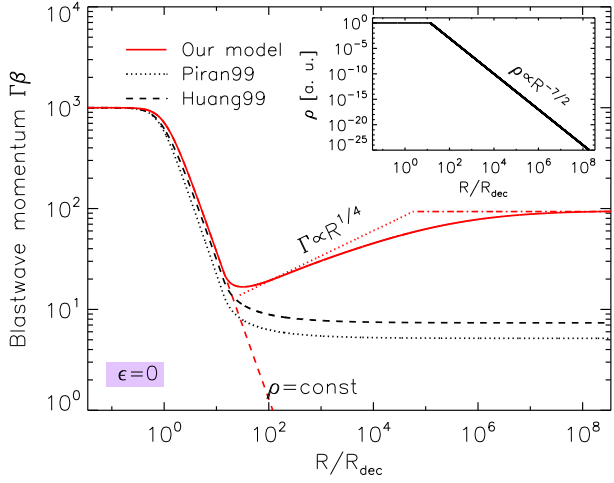}
\caption{
Evolution of the momentum $\Gamma\beta$ for an adiabatic blast wave propagating in a 
structured density profile, which passes from a uniform medium to a $\rho\propto R^{-7/2}$ 
profile at $R_b=15\,R_{dec}$. 
The solution of our model is shown with the red solid line, 
whereas the predictions by P99 and H99 are plotted with a dotted and a dashed line, respectively. 
It is apparent that the models by P99 and H99 cannot capture the re--acceleration of the blast 
wave occurring at $R\gtrsim 15\,R_{dec}$. We also show the solution corresponding to uniform density everywhere (red dashed line), the scaling $\Gamma\propto R^{1/4}$ predicted by \citet{shapiro80} for the re--acceleration phase (red dotted line), and the expected saturation at late times (red dot-dashed line).
}
\label{fig:reaccel}
\end{figure}
The formalism described in this paper is valid for any spherically symmetric
profile of the external density. 
If the blast wave encounters a change in the CBM profile during the deceleration phase, with density decreasing faster than $\rho\propto R^{-3}$ beyond some radius, the conversion of the accumulated internal energy back into bulk kinetic energy can lead to a re--acceleration of the blast wave. As shown by \citet{shapiro80}, the Lorentz factor of re--accelerating blast 
waves should scale as $\Gamma\propto R^{(s-3)/2}$.
Of course, the re--acceleration happens only for adiabatic blast waves; if the shell is fully radiative, all the internal energy is promptly radiated away, and it cannot produce any re--acceleration.

We consider the case of an adiabatic blast wave propagating in a medium where the external density is
constant up to $R_b=15 R_{dec}$ and then it decreases as $\rho\propto R^{-7/2}$. 
\fig{reaccel} shows the evolution of the shell momentum.
Our model (red solid line) shows that the evolution of the shell momentum departs 
from the result expected for a homogeneous medium (red dashed line) at radii $R\gtrsim R_b$, 
where the shell starts to re--accelerate. Its momentum increases as 
$\Gamma\propto R^{1/4}$ (red dotted line), as expected for a density profile with $s=7/2$. 
At large radii, the shell Lorentz factor asymptotes to a value which is lower than \go\ 
(we find $\Gamma\sim10^2$ in the example in \fig{reaccel}). 
This is just a consequence of energy conservation. 
Only the internal energy stored in the shell at the end of 
the deceleration phase ($R=R_b$) can be used to re--accelerate the blast wave. 
If $m_b\equiv m(R_b)$ is the mass accreted 
at $R_b$, the asymptotic value of the shell Lorentz factor can be obtained directly from the 
conservation of the bulk kinetic energy:
\be
\Gamma-1=(\Gamma_0-1)\frac{M_0}{M_0+\frac{9-2s}{17-4s} m_b}~~,
\ee
where the factor $(9-2s)/(17-4s)$ comes from the correction we adopt to 
recover the normalization of BM76 for adiabatic decelerating blast waves 
(see \eq{corr}). 
The capability to capture the blast wave re--acceleration is peculiar to our 
model and it requires a proper description of the adiabatic expansion losses. 
In contrast, the methods by P99 or H99 cannot describe the re--acceleration phase (dashed and dotted black curves in \fig{reaccel}, respectively). 

We find that the evolution of $\Gamma$ with radius in the re--accelerating phase departs 
from the scaling predicted by \citet{shapiro80} for large values of $s$, and it approaches 
$\Gamma\propto R^{3/4}$ in the limit $s\gg1$, 
which is the expected behavior for a fireball expanding in the vacuum, as we demonstrate in Appendix \ref{vacuum}. 

In summary, our model can be used to follow the dynamics of blast waves that propagate in a structured CBM, with density enhancements (or ``clumps'') and troughs, as it is often considered to explain flares and  rebrightenings in the optical and X-ray light curves of many GRB afterglows \citep{nakar_granot07,mesler12,hascoet12}. However, we remark that our results, being derived in the homogeneous shell approximation, 
cannot describe properly the role played by reverse shocks and rarefaction waves created 
when the shell interacts with a sudden density jump or drop. 
For this, one needs specific analytical treatments (e.g., \citealt{nakar_granot07}) 
or numerical simulations (e.g., \citealt{mimica_giannios11}). Also, in the present work we only consider a spherical shell, and we defer the discussion of jetted outflows (and their associated lateral expansion) to a follow--up paper (Sironi et al., in preparation).

\section{Results}

\subsection{Bolometric light curves}\label{lightcurve}

To predict the afterglow light curves we need to estimate how the observed luminosity 
evolves as a function of the observer's time. 
We only focus on the bolometric light curve, so we do not need to specify the
emission mechanism. We proceed in two steps: first, we estimate the luminosity as a function of the 
shock radius $R$, and then we introduce the relation between $R$ and the observer's time $t$. 

The luminosity observed from a parcel of shocked fluid moving at an angle $\theta$ 
with respect to the line of sight will be $L_{\theta}(\theta)=\delta^4 L'$, where 
$\delta=[\Gamma(1-\beta\cos\theta)]^{-1}$ is the Doppler factor of the shocked fluid, 
and $L'$ is the comoving luminosity. 
Assuming $L'$ to be isotropic in the comoving frame, the observed 
(isotropic equivalent)
luminosity $L$ is calculated by averaging $\delta^4$ over the angle $\theta$ \citep{ghirlanda12}:
\be
L=\frac{L'}{2}\int_0^\pi \delta^4 \sin \theta\, d\theta=\Gamma^2\frac{\beta^2+3}{3}L'
\ee
In our formalism the comoving luminosity is $L'=dE'_{rad}/dt'$, 
where $dE'_{rad}=\epsilon\,(\Gamma-1)\,dm\,c^2$ is the energy lost to radiation in the fluid comoving frame, and for spherical explosions $dm=4\pi R^2 \rho  dR$. 
By employing this expression for $dE'_{rad}$, we are implicitly saying that the emission at any given time only comes from the mass element $dm$ just passed through the shock. In other words, we are assuming that the electrons radiate right behind the shock, and then they evolve adiabatically. It follows that  the expression we find below for the observed luminosity is  valid only in the fast--cooling regime.

We then need to compute the distance $dR$ traveled by the shock during a comoving  
time interval $dt'$, which turns out to be
\be
\label{dtdr}
dR=\frac{\beta_{sh}}{\Gamma(1-\beta\beta_{sh})}c\,dt'~~,
\ee
where $\beta_{sh}$ is the shock velocity in the progenitor 
frame.\footnote{One could naively expect that $dR=\beta_{sh} \Gamma c dt'$. 
The difference with respect to the expression we employ comes from the fact that, 
in our approach, a given fluid element radiates only right after passing through 
the shock, so two subsequent emitting events are spatially coincident in the shock 
frame, rather than in the fluid comoving frame (as implicit in $dR=\beta_{sh} \Gamma c dt'$).}
This expression can be simplified by employing the convenient parameterization 
$\beta_{sh}=4\Gamma^2\beta/(4\Gamma^2-1)$, which is valid both in the ultra--relativistic 
phase (giving $\Gamma_{sh}^2\simeq2\Gamma^2$) and in the non--relativistic limit 
($\beta_{sh}\simeq4/3\,\beta$). In this relation for $\beta_{sh}$, the speed $\beta$ 
should be interpreted as the fluid velocity right behind the shock, before the radiative 
cooling losses have appreciably changed the dynamics of the post--shock flow 
\citep[see][]{cohen98}. 
It follows that $(1-\beta\beta_{sh})^{-1}=\Gamma^2(\beta^2+3)/3$, 
and the comoving luminosity is
\be
L'=\epsilon (\Gamma-1)\Gamma\,\frac{\beta^2+3}{3}\beta_{sh}\frac{dm}{dR} c^3~~,
\ee
so that the observed bolometric luminosity as a function of the shock radius will be
\be
\label{eq:lumin}
L=\epsilon (\Gamma-1)\Gamma^3\,\left(\frac{\beta^2+3}{3}\right)^2\beta_{sh}\frac{dm}{dR} c^3~~.
\ee
The linear dependence on the shock velocity is expected, since in our model the luminosity 
is proportional to the rate at which the shock picks up matter from the circumburst medium. 
In the non--relativistic limit, the observed luminosity reduces to 
$L=\epsilon(1/2\,\beta^2)\,\beta_{sh}\,dm/dR \,c^3$, as expected from a shock with 
velocity $\beta_{sh}$ that converts a fraction $\epsilon$ of the post--shock internal 
energy per unit mass $(1/2)\, \beta^2 c^2$ into radiation.

Finally, to obtain the luminosity as a function of the observer's time, 
we need to relate the shock radius $R$ to the observer's time $t$. 
For the sake of simplicity, we do not consider the formalism of equitemporal 
surfaces discussed by \citet{bianco_ruffini05}, but following \citet{waxman97}, 
we use the relation
\be
\label{eq:time}
t(R)=t_{R}+t_{\theta}=\int_0^R{\frac{1-\beta_{sh}}{\beta_{sh} c}dr}+\frac{R}{\Gamma^2(1+\beta)c}
\ee
which consists of the sum of a radial delay $t_{R}$, i.e., the difference between 
light travel time to radius $R$ and shock expansion time to the same radius, and an 
angular delay $t_{\theta}$, for photons emitted from the same radius $R$ but at 
different angles with respect to the line of sight. 
The angular term $t_\theta$ dominates in the ultra--relativistic limit, 
whereas the radial term $t_{R}$ prevails in the non--relativistic phase.

We point out that the expression for the observed luminosity in \eq{lumin} and the relation 
$t(R)$ in \eq{time} are valid both in the ultra--relativistic and non--relativistic phases. 
From their combination, one could estimate the asymptotic scaling of the observed luminosity 
with time, as the blast wave decelerates. 
For adiabatic blast waves, we find $L\propto \epsilon(t)E_0\, t^{-1}$ both in the 
relativistic and \nr\ regimes (provided $\epsilon(t)\ll1$ at all times). Note that in both regimes the bolometric luminosity does not depend on the value of the CBM density.
For radiative blast waves we find $L\propto [M_0^{8-2s} t^{-(10-3s)}/A_0]^{1/(7-2s)}$ 
in the relativistic phase and $L\propto [M_0^{5-s}t^{-(7-2s)}/A_0]^{1/(4-s)}$ 
in the \nr\ limit (assuming $\epsilon(t)\sim1$ at all times).

\begin{figure}
\vskip -0.5 cm
\hskip -0.5 truecm
\includegraphics[scale=0.52]{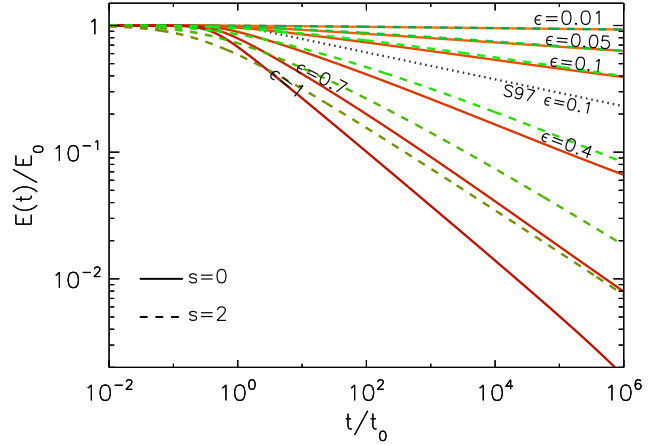}
\vskip -0.3 cm
\caption{
Blast wave energy content (normalized to its initial value) as a function of the observer's time for a 
\bw\ with constant efficiency $\epsilon$. Different values of $\epsilon$ (from 0.01 to 1) 
are considered, both for $s=0$ (solid lines) and $s=2$ (dashed lines). 
For comparison, the solution proposed by \citet{sari97} for $\epsilon=0.1$ and $s=0$ is 
also shown (dotted black line). 
}
\label{fig:evo_time}
\end{figure}

\begin{figure}
\hskip -0.3 truecm
\includegraphics[scale=0.5]{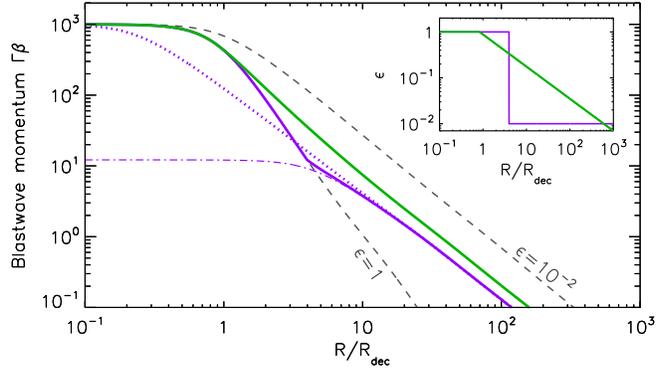}
\caption{
Blast wave momentum $\Gamma\beta$ as a function of $R$ for a shell with radiative 
efficiency $\epsilon$ that evolves in radius as shown in the inset. 
Two different cases for $\epsilon(R)$ are considered (purple and green lines). 
The dashed grey curves refer to the standard adiabatic and radiative cases with 
$\epsilon$ constant and equal to $10^{-2}$ and 1, respectively. 
The dotted and dot--dashed purple curves show the adiabatic  
solutions with $\epsilon=const=10^{-2}$ having the same late--time energy as the purple solid curve, and either the same $\Gamma_0$ (purple dotted line) or the same $M_0$ (purple dot-dashed line).
}
\label{fig:epsilon_r}
\end{figure}

\subsection{From the radiative to the adiabatic regime}
\label{epsilon}
One of the crucial parameters that could help to unveil the nature of the 
progenitor of GRBs is the total energy output of the central engine $E_{tot}$.
This quantity can be estimated from the sum of the energy emitted during the prompt phase 
($E_\gamma$) and the energy of the blast wave that powers the afterglow emission ($E_0$). Moreover, from these two quantities one can derive the efficiency of energy dissipation 
in the prompt phase $\eta=E_\gamma/E_{tot}$, which could help to 
constrain the uncertain mechanism that produces the prompt $\gamma$-rays.

Estimates of $E_0$ require observations of the afterglow over a broad range in time and 
frequency (e.g., \citealt{panaitescu02,yost03}). 
An alternative method is based on
observations at frequencies above the synchrotron cooling frequency 
$\nu_{c}$, since they give a robust proxy for the total blast wave energy $E_0$
\citep{kumar00,freedman01}. 
For typical parameters, the X--ray data at late times lie 
above $\nu_{c}$, so the X--ray flux can be used to constrain the blast wave energy  
 $E_0$ and the efficiency of the prompt emission $\eta=E_\gamma/(E_\gamma+E_0)$.
Several authors have applied this method to different samples of both long and short  
GRBs \citep{lloyd-ronning04,zhang07,racusin11,berger07}, finding similar results: $\eta$ 
reaches values of 10\% or higher, challenging 
the simplest version of the internal shock model, where only a few per 
cent of the jet energy can be dissipated \citep{kumar99,panaitescu99}. 
{\it Swift} observations make the situation even worse, raising some of the inferred GRB 
prompt efficiencies up to $>$90\% \citep{zhang07}.

The problem can be alleviated if the afterglow energetics is underestimated.
The adiabatic model cannot be strictly correct, since radiation takes away some energy from the blast wave.
The X--ray flux is a proxy for the energy content of the \bw\ at late times, but this is smaller than the initial energy  $E_0$, if the evolution is not perfectly adiabatic.
By neglecting the radiative losses, one may underestimate $E_0$ and then overestimate $\eta$. 
Radiative corrections based on the equation proposed by \citet{sari97} have been considered 
by \citet{lloyd-ronning04}.
Even in this case, the inferred efficiencies $\eta$ are relatively high (between 40\% and 100\%).
The validity of the equation derived by \citet{sari97} has been questioned, since it 
employs the relation $R=16\Gamma^2ct$ between shock radius and observer's time, instead of the correct expression in Eq.~\ref{eq:time} \citep[see][]{waxman97}.
So, the importance of radiative corrections in the afterglow energetics at late times is still unclear. In Fig. \ref{fig:evo_time}, we use 
our model to estimate, for constant values of $\epsilon$ ranging from $10^{-2}$ to 1, the energy 
content of the fireball $E(t)$ as a function of the observer's time, normalized 
to $t_0=R_{dec}/(\Gamma_0^2c)$.
As an example, for a fireball with  $E_0=10^{53}\,\rm{ergs}$ and $\Gamma_0=10^3$ propagating in a medium with uniform density $n_0=1\,\rm{cm^{-3}}$, the deceleration time is $t_0\sim1$s. The plot in  Fig. \ref{fig:evo_time} shows that, for a constant radiative efficiency $\epsilon=0.1$, the energy content of the blast wave decreases by a factor of two in the first $10^4$ seconds. For higher values of $\epsilon$, the fraction of fireball energy lost to radiation will be even larger.
For comparison, we also plot the solution proposed by \citet{sari97} for $\epsilon=0.1$, 
which tends to overestimate the radiative losses.  

The results presented in Fig. \ref{fig:evo_time} rely on the assumption that the 
efficiency $\epsilon$ remains constant in time. The results significantly change if GRB afterglows experience an early fully-radiative phase followed by the standard adiabatic evolution at later times.  
\begin{figure}
\hskip -0.3 truecm
\includegraphics[scale=0.5]{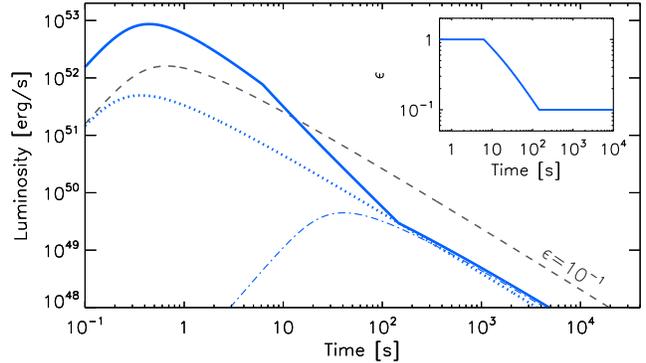}
\caption{
Solid line: light curve of a blast wave with time--dependent $\epsilon$. 
The evolution of $\epsilon(t)$ is shown in the inset. 
The grey dashed line shows the light curve of a blast wave with the same values of 
$E_0$ and \go, but $\epsilon=const$ and equal to the value reached by the
time-dependent $\epsilon(t)$ at late times (i.e., $\epsilon=0.1$). 
Dot--dashed and dotted lines show the light curves of two \bw s with constant $\epsilon=0.1$, 
that have an initial energy equal to the late time energy content of the \bw\ 
with time-dependent $\epsilon$. 
In particular, the dotted curve refers to the case of identical initial \lf\  $\Gamma_0$, while the 
dot--dashed line refers to the case of identical initial mass $M_0$.
}
\label{fig:epsilon_t}
\end{figure}
To address this case, in \fig{epsilon_r} we study the evolution of the blast wave momentum $\Gamma\beta$ when 
$\epsilon$ decreases from $\epsilon=1$ to $\epsilon=10^{-2}$ (green and purple solid lines), and 
we compare it to the cases with constant $\epsilon=1$ and constant $\epsilon=10^{-2}$ 
(dashed lines). 
We consider two possible choices for the time evolution of $\epsilon$, as shown in the inset of \fig{epsilon_r}: 
(\textit{i}) a step function where $\epsilon=1$ up to a few $R_{dec}$ and $\epsilon=10^{-2}$ at larger radii (purple line); or (\textit{ii}) $\epsilon=1$ for $R\leq R_{dec}$ 
and $\epsilon\propto R^{-0.7}$ for $R>R_{dec}$ (green line). 
In the latter case (green solid line), the blast wave 
momentum $\Gamma\beta$ initially follows the radiative solution (dashed line with $\epsilon=const=1$), and 
then smoothly approaches the scaling $R^{-3/2}$ expected for
adiabatic blast waves.  
At large radii, the difference in normalization with respect to the solution having $\epsilon=const=10^{-2}$ (dashed line) comes from the fact that the fireball with evolving $\epsilon$ 
has lost some of its energy during the early radiative phase. 

Consider now the case of a step function for the evolution of $\epsilon(R)$ (purple solid line in \fig{epsilon_r}). 
The \bw\ momentum initially follows the radiative solution, but
at radii larger than $4R_{dec}$ we have $\epsilon(R)=10^{-2}$ and the energy content remains nearly constant 
$E(R>4R_{dec})\sim const=E_{0,f}$. 
At large radii, the blast wave evolves as a 
fireball with $\epsilon=const=10^{-2}$ and initial energy $E_0=E_{0,f}$. 
This is demonstrated in \fig{epsilon_r} for two different cases: (\textit{i}) a fireball with $E_0=E_{0,f}$ 
and the same ejecta mass $M_0$ as the shell with evolving 
$\epsilon$ (purple dot--dashed curve); and (\textit{ii}) the case of a fireball with $E_0=E_{0,f}$ 
and the same initial \lf\ \go\ as the shell with evolving $\epsilon$ 
(purple dotted curve).

Let us now compute the bolometric light curve in the case of a time-evolving $\epsilon$. 
We are interested in reproducing the observations at GeV energies in the scenario of an early radiative phase of the GRB external shock. At the same time, we want to be consistent with the standard adiabatic evolution at later times.
The light curve in the  initial radiative regime (lasting a few tens of seconds) will be characterized by a steep temporal decay 
 similar to that observed in the GeV emission \citep{ghisellini10}, which will be followed by the standard  adiabatic scaling $L\propto t^{-1}$  
seen in the X--ray and optical bands. 
In this simple scenario we do not attempt to explain the 
plateaus and flares often present in the X--ray and/or optical data. 
Although bolometric light curves cannot be directly compared to the observations in a given frequency band, they are still useful to estimate how much energy can be radiated and lost by the 
blast wave before the adiabatic regime sets in. 
Consider the evolution of $\epsilon$ shown in the inset of \fig{epsilon_t}. At early times, the blast wave is fully radiative (i.e., $\epsilon(t)=1$). 
After a few seconds, the radiative efficiency decreases, reaching at $t\sim10^2\,$s  
the final value $\epsilon=0.1$, which is the typical value inferred from late 
afterglow data. The resulting light curve is shown with the solid curve 
in \fig{epsilon_t}. 
We also present, for comparison, the light curve of a fireball with the same initial 
values of $E_0$ and \go, but with $\epsilon=const=0.1$ 
(dashed grey line). 
After $10^2$ s, the two light curves have the same slope $L\propto t^{-1}$, and the 
difference in their normalizations gives the ratio between the initial energy $E_0$ 
and the late-time energy content of the \fb\ with time-varying $\epsilon$. 
In the case shown in  \fig{epsilon_t}, this ratio is $\sim$6, implying that $\sim83\%$ of 
the initial energy $E_0$ has been radiated in the first $10^2$ seconds. 
By using observations after $10^2\,$s, as it is usually the case for optical and X-ray data, one would 
underestimate the blast wave energy $E_0$, and so 
overestimate  the prompt efficiency $\eta$. 
Broadband fits of GeV, X--ray and optical data are needed in order to understand 
if an early radiative phase is required to fit the GeV light curves and, if this is the case, to estimate how much energy is radiated before the afterglow brightens in the X-ray and optical bands
(Nava et al., in preparation).

\subsection{Estimate of the initial bulk Lorentz factor \go} \label{gamma0}
During the initial coasting phase, when the swept--up matter is negligible and the shell \lf\ is constant, 
the bolometric luminosity scales with time as $t^{2-s}$ (i.e., it is rising for $s<2$). 
When the deceleration starts (approximately at $R_{dec}$), the \lf\ decreases, producing 
a feature in the light curve: a peak (if $s<2$) or a change in the power--law decay slope (if $s>2$). 
The transition between the two regimes occurs approximately at 
$t_0=R_{dec}/(\Gamma_0^2c)$ \citep{rees92}, where $R_{dec}$ has been defined
in \eq{rdec}.
Since $t_0$ depends on the initial \lf\ (and it is nearly insensitive to all other parameters),
the measurement of this time can be used to infer \go\ \citep{rees92,sari99}.
Since the peak time $t_{peak}$ of the afterglow light curve is a good proxy for $t_0$, we focus in this section on the relation between the peak time and \go.
We consider both adiabatic and semi--radiative regimes.

\vskip 0.2 cm
\noindent
{\it Adiabatic case ---}
Both before and after the peak time $t_{peak}$, the luminosity scales in time as a power law. 
From \eq{lumin}, by setting $\Gamma=\Gamma_0$ we derive $L(t<t_{peak})$,
while using $\Gamma(R)$ given by the BM76 solution we derive $L(t>t_{peak})$ 
(in both epochs we also use the relativistic approximation $\beta\sim\beta_{sh}\sim1$).
Following \citet{ghirlanda12}, we extrapolate the two power-law segments of $L(t)$ and find 
their intersection time, $t_{\wedge}$. 
The initial bulk Lorentz factor \go\ is related to the intersection time $t_{\wedge}$ by the equation:
\begin{equation}\label{eq:g0_ad}
\Gamma_0(\epsilon=0,s)=\left[\frac{(17-4s)(9-2s)3^{2-s}}{2^{10-2s}(4-s)
\pi\, c^{5-s}}\right]^\frac{1}{8-2s} 
\left(\frac{E_0}{A_0t_{\wedge}^{3-s}}\right)^\frac{1}{8-2s}~~.
\end{equation}
Even if $t_\wedge$ does not coincide with $t_{peak}$, for adiabatic models and $s=0$ 
it is possible to  find numerically that $t_\wedge=0.9\,t_{peak}$.
Based on our model, the relation between $t_{peak}$ and \go\ 
for an adiabatic blast wave expanding in a homogeneous medium is then:
\begin{equation}\label{eq:g0_st}
\Gamma_0(\epsilon=0,s=0)=190\left[\frac{E_{\gamma,53}}{n_0\,t_{peak,2}^3\,\eta_{0.2}}\right]^{1/8}
\end{equation}
where $n_0$ is defined by $\rho=m_pn_0R^{-s}$, $E_{\gamma,53}=E_{\gamma}/10^{53}\rm ergs$, $\eta_{0.2}=\eta/0.2$, and $t_{peak,2}=t_{peak}/100\rm \,s$. We have replaced $E_0$ by considering the relation $E_0\simeq E_\gamma/\eta$, where $E_\gamma$ is the energy emitted in the prompt phase and $\eta$ is the radiative efficiency of internal shocks. 
We now compare our formula with those proposed in the literature for the 
adiabatic regime and $s=0$. 
The equations most frequently used to estimate \go\ from $t_{peak}$ 
are those derived by \citealt{molinari07} (M07 hereafter) 
\citep{liang10,gruber11,rykoff09,evans11} and by \citealt{sari99} (SP99). 
SP99 suggest the formula:
\begin{equation}
\Gamma_0=\left( \frac{3E_0}{32\pi A_0 c^5 t_{peak}^3}\right)^{1/8}=160\left[\frac{E_{\gamma,53}}{n_0\,t_{peak,2}^3\,\eta_{0.2}}\right]^{1/8}
\end{equation}
while for M07 this expression represents $\Gamma_{dec}$ (the \lf\ at the deceleration 
radius $R_{dec}$), and to derive \go\ they arbitrarily assume $\Gamma_0=2\Gamma_{dec}$.
Note that our formula differs from the existing ones only by a numerical factor, since the dependences 
of \go\ on $E_0$, $A_0$, and $t_{peak}$ are exactly the same.
We estimate both numerically and analytically the discrepancies between our 
formula and those proposed by M07 and SP99.
We find that our estimate of \go\ is a factor of 1.64 systematically lower than the estimate by M07 and 
a factor of 1.20 larger than the estimate by SP99.
These numbers are quite insensitive to the choice of $E_0$ and $A_0$.

\begin{figure}
\hskip -0.3 truecm
\includegraphics[scale=0.45]{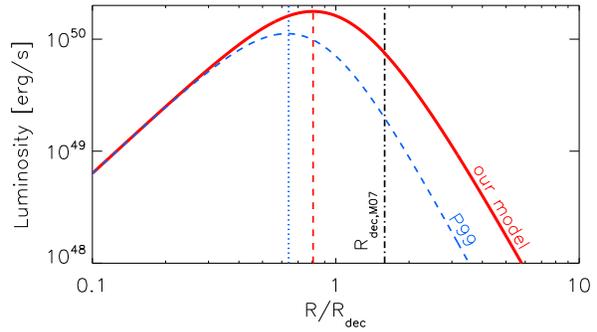}
\caption{
Light curve as a function of $R/R_{dec}$ (where $R_{dec}$ is defined as in Eq.~\ref{eq:rdec}). 
The red solid curve is obtained by using our model, while the blue dashed curve is based on the P99 model. 
The vertical dashed and dotted lines mark the peak of the light curves for the two models. 
The dot-dashed line marks, instead, the deceleration radius as defined by M07.
}
\label{fig:l_r}
\end{figure}

We conclude that the equation proposed by M07 leads to a significant overestimate of 
\go\ (by a factor of 1.64), while the equation presented by SP99 gives a lower estimate of \go, 
in better agreement with our result.
The main source of disagreement between our model and the approach by M07 is 
in their definition of the deceleration radius.
In their model, the peak time \tpeak\ of the light curve is assumed to correspond to the deceleration 
time,  when the fireball reaches the deceleration radius.
While SP99 define the deceleration radius as the radius where the swept--up matter is
$m($\rdec$)=M_0/$\go$=E_0/(\Gamma^2_0c^2)$ (i.e., the same definition adopted in this work), 
M07 use the relation $m(R_{dec,M07})=E_0/(\Gamma_{dec}^2c^2)$ (see also \citealt{lu12}). 
Fig. \ref{fig:l_r} shows that the luminosity peaks close to $R_{dec}$, 
which is then a good proxy for the onset of the deceleration epoch. 
The deceleration radius $R_{dec,M07}$ defined by M07, instead, 
overestimates the radius at which the luminosity reaches its maximum by a factor of $\sim2$.
This difference becomes even larger when 
considering the light curve derived with the P99 model (dashed blue line).

\vskip 0.2 cm
\noindent
{\it Semi--radiative case ---}
The same approach can be applied to generic values of $\epsilon$, but in this 
case it is not possible to find an analytic formula that 
describes the \lf\ after the peak time (and then an analytic equation for $L(t>t_{peak})$). 
We then define:
\begin{eqnarray}\label{eq:goeps}
\Gamma_0(\epsilon,s)=K(\epsilon,s) \left(\frac{E_0}{n_0t_{\wedge}^{3-s}}\right)^\frac{1}{8-2s}
=K^\prime(\epsilon,s)\left(\frac{E_0}{n_0t_{peak}^{3-s}}\right)^\frac{1}{8-2s}
\end{eqnarray}
and we list in Tab.~\ref{tab:g0} the values of $K$ and $K^\prime$ numerically estimated for 
different choices of $\epsilon$ and $s$.

\begin{table}
 \centering
 \begin{minipage}{80mm}
  \caption{Normalization factors for the estimate of \go\ as a function of $t_\wedge$ ($K$) and $t_{peak}$ ($K^\prime$), calculated for different values of $\epsilon$ and $s$ (see \eq{goeps}).}
  \begin{tabular}{@{}clcccc@{}}
  \hline
   Density     &  &$\epsilon<10^{-2}$  &  $\epsilon=0.1$  &  $\epsilon=0.5$ & $\epsilon=1$\\
   profile & & & & &\\
 \hline
 
 s=0 &   $K$             &1.95$\times 10^{-4}$ & 1.94$\times 10^{-4}$ & 1.86$\times 10^{-4}$& 1.74$\times 10^{-4}$\\
        &  $K^\prime$  &2.08$\times 10^{-4}$ & 2.03$\times 10^{-4}$ & 1.88$\times 10^{-4}$& 1.75$\times 10^{-4}$\\
        \hline
 s=2 &   $K$             &6.70$\times 10^{-3}$ & 6.60$\times 10^{-3}$ & 5.90$\times 10^{-3}$& 4.95$\times 10^{-3}$\\
 \hline
 \label{tab:g0}
\end{tabular}
\end{minipage}
\end{table}

\section{Conclusions}
We consider the interaction of a spherical relativistic \bw\ with the surrounding medium
and we describe its dynamical evolution by taking into account radiative  and adiabatic losses.
We derive the equation that governs the evolution of the \bw\  Lorentz factor as a function of the shock radius (\eq{G solve}). 
This equation is general and independent of the composition of the external 
medium and of the nature and efficiency of the emission processes. 
To present a simple solution of this equation, however, we restrict ourselves to the case where radiative losses are important only for the freshly shocked matter (which is strictly true only in the fast cooling regime), or to the case where radiative losses are not important at all (quasi-adiabatic regime). 
The equations we derive are valid for any generic (spherically symmetric) density profile.
We also give the correction factor that should be employed, in our homogeneous thin shell approximation, 
to match the Blandford \& McKee (1976) solution in the relativistic deceleration stage.

\begin{itemize}
\item {\it Radiative losses ---} Our model is valid for any value of the overall radiative efficiency $\epsilon=\epsilon_e\epsilon_{rad}$, where \epse , as usual, is the fraction of shock-dissipated energy gained by the electrons, while \epsrad\ is the fraction of this energy which is lost to radiation. The value of $\epsilon$ can evolve with time, thus varying the impact of the radiative energy losses on the shell dynamics. The decrease in the blast wave energy content due to the radiative losses is taken into account self-consistently in our approach. 

We study models where the value of $\epsilon$ is kept constant in time and we derive, for each value of $\epsilon$, the evolution with time of the energy content of the blast wave (\fig{evo_time}). For $\epsilon$$\,\gtrsim\,$0.1, the radiative losses can decrease the energy content of the fireball within $t\sim10^4\,$s by more than a factor of two. In the extreme limit $\epsilon$$\,\simeq$1, the energy content decreases by an order of magnitude in the first $\sim$10$^2\,$s. As shown in \fig{slopes}, radiative losses change the slope $\alpha$ of the scaling in radius of the blast wave Lorentz factor (such that $\Gamma(R)\propto R^{-\alpha}$), from the classical $\alpha=(3-s)/2$ of the adiabatic BM76 solution to the steeper $\alpha=(3-s)$ of the fully radiative solution. 

We have also considered the realistic case where $\epsilon$ varies in time from $\epsilon\simeq1$ (fully-radiative) down to $\epsilon\ll1$ (quasi-adiabatic). The corresponding blast wave Lorentz factor is shown in \fig{epsilon_r}, and the expected bolometric light curve is presented in \fig{epsilon_t}. In this case, the luminosity after the peak initially decreases as a steep function of time, and then it follows a flatter decay.\\

\item {\it Adiabatic losses ---} Both electrons and protons suffer from adiabatic losses due to the fireball expansion. Adiabatic losses describe the re--conversion of the internal random energy generated at the shock into bulk kinetic energy. A proper treatment of the adiabatic losses is important in two ways. First, when the velocities are \nr\ we can recover the correct scaling $\beta\propto R^{-3/2}$ between the fluid velocity and the shock radius predicted by the Sedov--Taylor self--similar solution. Models that do not account for the adiabatic losses (P99) predict instead a steeper scaling, as $\beta\propto R^{-3}$ (see the dotted line in the grey area of \fig{momentum}). 
Second, a proper description of the adiabatic losses allows to capture the blast-wave re--acceleration in the case  of an ambient medium where the  density decreases faster than $R^{-3}$, which cannot be properly described neither by P99 nor by H99 (\fig{reaccel}). 

\end{itemize}

The model we present is particularly suitable to study the initial expansion of GRB external shocks, 
when a peak (or a change of slope) in the light curve is expected to arise. 
We propose an equation to derive the initial Lorentz factor \go\ from the peak time of the light curve. Our formula is  valid for any value of $\epsilon$ and $s$ 
(see Eq.~\ref{eq:goeps} and the normalization factors listed in Tab.~\ref{tab:g0}).
In the adiabatic ($\epsilon<10^{-2}$) and homogeneous ($s=0$) case we compare our formula in \eq{g0_st} with the equations proposed in the existing literature (e.g., M07 and SP99, which differ from each other by a factor of 2). 
We predict a value for \go\ which is intermediate between M07 and SP99, yet showing a better agreement with the formula of SP99. 
The equation  by M07, often adopted in the literature, tends to overestimate \go\ by a factor of $\sim1.6$. As an example, for the two GRBs studied in M07, we find \go$\sim250$ instead of \go$\sim400$. The formula derived in this paper is consistent with that used in Ghirlanda et al. 2012 for the estimate of \go\ for GRBs with a peak in their optical light curve.

However, some caveats should be pointed out. Our estimate of $\Gamma_0$, similarly to the formulae by SP99 and M07, is based on the bolometric light curve, while the observations are performed in a given frequency band. Moreover, our estimate for the photon arrival time (see \eq{time}, which is used to compute the luminosity) relies on the simplifying assumption of emission from a spherical shell. Effects related to the shape of the  radiation spectrum, the  geometry  of the emitting material and the viewing angle (in the case of a collimated outflow) can affect the relation between the peak time of the light curve observed at a given frequency and \go. However, we point out that the approach by M07 (commonly adopted in the literature) tends to overestimate \go , independently from the caveats listed above, since the difference mostly arises from their overestimate of the deceleration radius, as shown in \fig{l_r}.

A complete modeling of the afterglow flux in a given frequency band cannot be presented here, since it requires a description of  the radiative processes responsible for the observed emission, which we have not considered in this work. 
The dynamical model presented here coupled to a radiative model accounting for synchrotron and inverse Compton losses will allow us to compute spectra and light curves at different frequencies and to investigate several issues related to the afterglow physics. 
Moreover, the introduction of a radiative model will allow to
overcome some of the simplifications adopted in this work, such as the requirement of fast cooling, implicitly assumed in deriving the formula for the luminosity in \S4.1.
In particular, we suggest that our dynamical model will be particularly suitable to study:
\begin{itemize}
\item the relation between GeV emission and external shocks. In a forthcoming paper, we plan to investigate whether the high--energy GeV emission in GRBs can originate from afterglow radiation in external shocks. In particular, thanks to the formalism developed in this work, we will be able to account for the presence of electron-positron pairs, the role of radiative losses, and the temporal evolution of the radiative efficiency from a highly radiative phase to the standard adiabatic regime.
\item complex circum-burst density profiles and their effect on the light curve, in order to explain the rebrightenings, plateaus and bumps sometimes detected in the optical and X--ray data.

\end{itemize}

\appendix
\section{Re--acceleration into vacuum}
\label{vacuum}

We show that the Lorentz factor of an adiabatic blast wave scales as $\Gamma\propto R^{3/4}$ 
as the shock reaccelerates into the vacuum. 
We assume that the circum burst mass density has some profile $\rho(R)$ up to a radius $R_{vac}$ 
(where the blast wave is still ultra--relativistic), and after that the density drops to zero. 
For $R>R_{vac}$, Eq. \ref{eq:G solve} reduces to
\be
\label{eq:Gvac solve}
\frac{d\Gamma}{dR}=-\frac{\Gamma_{\it eff}
\frac{dE'_{ad}}{dR}}{(M_0+m_{vac})\,c^2+E'_{int}\frac{d\Gamma_{\it eff}}{d\Gamma}}
\ee
where $m_{vac}=4 \pi\int_0^{R_{vac}} \rho(r) r^2 dr$ is the swept--up mass up to radius $R_{vac}$. 
For an ultra--relativistic adiabatic blast wave in the deceleration phase, the comoving 
internal energy satisfies $E'_{int}\gg (M_0+m) c^2$, and the same will hold at the early 
stages of the subsequent re--acceleration phase. 
In addition, we assume that for $R\gtrsim R_{vac}$, most of the post--shock plasma is still 
highly relativistic (i.e., $\gamma'_{ad}\simeq p'_{ad}\gg1$), despite the adiabatic losses. 
From Eqs.~\ref{eq:int energy} and \ref{eq:ad losses}, we obtain
\be
\frac{d E'_{ad}}{dR}&=&-E'_{int}(R)\left(\frac{1}{R}-\frac{1}{3}\frac{d \log \Gamma}{dR}\right)~,
\ee
which relies on the assumption that the comoving shell thickness is $\sim R/\Gamma$ also in the 
re--acceleration phase, so that the comoving volume is $V'\propto R^3/\Gamma$ 
\citep[see][]{shapiro80}.
In the limit $E'_{int}\gg (M_0+m_{vac})\, c^2$, \eq{Gvac solve} then reduces to 
\be
\frac{d\Gamma}{dR}=\Gamma\left(\frac{1}{R}-\frac{1}{3}\frac{d \log \Gamma}{dR}\right)~,
\ee
whose solution is $\Gamma\propto R^{3/4}$.

One can wonder why we obtain $\Gamma\propto R^{3/4}$, while the
initial acceleration of the fireball follows $\Gamma\propto R$ 
\citep[e.g.,][]{meszaros06}.
The difference is due to different choices for the comoving volume $V'$.
During the initial acceleration phase $V'\propto R^3$
(i.e., a filled sphere), while in the case we are considering here
(i.e., forward shocks giving rise to the afterglow) we
assume $V'\propto R^3/\Gamma$, which corresponds to a shell of thickness $\sim R/\Gamma^2$ in the progenitor frame. If we were to assume $V'\propto R^3$, we would recover
$\Gamma\propto R$.

\section{Reverse shock}
\label{reverse shock}
The description of the reverse shock (RS) can be easily included in our model. When the RS is considered, four regions can be identified (see Fig.~B1): the unshocked CBM (region 1), the shocked CBM (region 2), the shocked ejecta (region 3) and the unshocked ejecta (region 4). We assume that the unshocked CBM is at rest in the progenitor frame and that regions 3 and 2 (the blast wave) move with the same bulk \lf, which is a reasonable assumption \citep{beloborodov06}.
\begin{figure}
\includegraphics[scale=0.33]{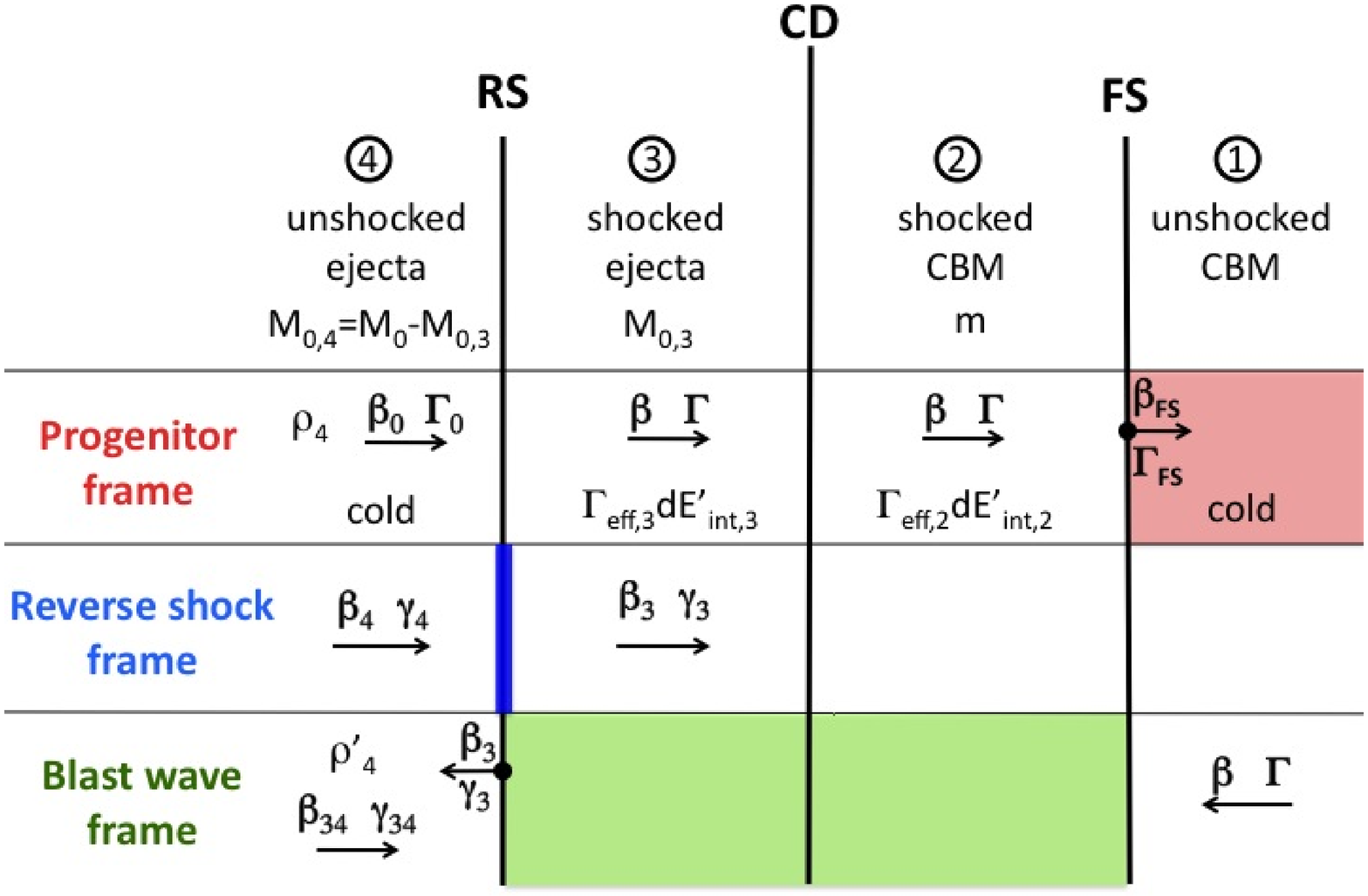}
\label{sketch}
\vskip -0.8 truecm
\caption{Sketch of the model considered when the reverse shock is included in the dynamics. In this case, four regions can be identified, separated by the reverse shock (RS), the contact discontinuity (CD) and the forward shock (FS). Each row shows the notation adopted in different frames, highlighted by the shaded regions: the progenitor frame (which coincides with region 1), the RS frame and the blast wave frame (regions 2 and 3).}
\end{figure}
Regions 2 and 3, however, can be characterized by different adiabatic indices, which we call $\hat\gamma_2$ and $\hat\gamma_3$. This implies that their effective Lorentz factors (defined according to Eq. \ref{eq:gammaeff}) may also be different.
Following the notation in Fig.~B1, the total energy in the progenitor frame now becomes:
\be
E_{tot}=\Gamma_0M_{0,4}\,c^2+\Gamma M_{0,3}\,c^2+\Gamma m\,c^2+\Gamma_{\it eff,3} E'_{int,3}+\Gamma_{\it eff,2} E'_{int,2}
\label{eq:EtotRS}
\ee
As before, this total energy changes due to the rest mass energy $dm\,c^2$ accreted from the CBM and due to the energy lost to radiation by the shocked CBM. In addition, in the scenario considered in this section, energy losses from the fraction of the ejecta shocked by the RS should also be considered:
\be
dE_{tot}= dm\,c^2+\Gamma_{\it eff,2} dE'_{rad,2}+\Gamma_{\it eff,3} dE'_{rad,3}
\ee
In analogy with the description we adopted for the shocked CBM, we use the following equations for the shocked ejecta:
\be
dE'_{int,3} = dE'_{sh,3}+dE'_{ad,3}+dE'_{rad,3}
\ee
\be
dE'_{sh,3}=(\gamma_{34}-1)dM_{0,3}\,c^2
\label{eq:esh3}
\ee
Here $\gamma_{34}$ is the Lorentz factor of the unshocked ejecta relative to the blast wave (see Fig.~B1), which can be expressed in terms of $\Gamma$ and $\Gamma_0$ (which is the Lorentz factor of the unshocked ejecta in the progenitor frame):
\be
\gamma_{34}=\Gamma_0\Gamma(1-\beta_0\beta)
\ee

Using the equations from \ref{eq:EtotRS} to \ref{eq:esh3} and $dM_{0,4}=-dM_{0,3}$, we can generalize Eq. \ref{eq:G solve}, the formal equation giving $\Gamma(R)$:

\begin{eqnarray}
\frac{d\Gamma}{dR}&=&-\frac{(\Gamma_{eff,2}+1)(\Gamma-1)\frac{dm\,c^2}{dR}+\Gamma_{eff,2}dE'_{ad,2}}{(M_{0,3}+m)c^2+E'_{int,2}\frac{d\Gamma_{eff,2}}{d\Gamma}+E'_{int,3}\frac{d\Gamma_{eff,3}}{d\Gamma}}+\nonumber \\
&&-\frac{(\Gamma-\Gamma_0-\Gamma_{eff,3}+\Gamma_{eff,3}\gamma_{34})\frac{dM_{0,3}c^2}{dR}+\Gamma_{eff,3}dE'_{ad,3}}{(M_{0,3}+m)c^2+E'_{int,2}\frac{d\Gamma_{eff,2}}{d\Gamma}+E'_{int,3}\frac{d\Gamma_{eff,3}}{d\Gamma}}
\end{eqnarray}

As before, we need to specify $dE'_{ad}$ and $E'_{int}$, but now we have adiabatic losses and internal energy both in region 2 and in region 3. In addition, we need to derive $dM_{0,3}$.

\vskip 0.2 cm
\noindent
{\it Derivation of $dM_{0,3}$} -- 
To derive the mass $dM_{0,3}$ that crosses the RS in a given time interval (corresponding to the time during which the FS travels a distance $dR$) we consider the frame at rest with the shocked ejecta (region 3):
\be
dM_{0,3}=4\pi R^2_{RS}(\beta_{34}+\beta_3)\,\rho'_4\,c\,dt'
\ee
where $\rho'_4$ is the mass density of the unshocked ejecta measured in the frame of the shocked ejecta. 
If $\rho''_4$ is the density measured in the frame 4 (i.e., the proper density) and $\rho_4$ is the density in the progenitor frame (see Fig.~B1), then:
\be
\rho'_4=\gamma_{34}\,\rho''_4=\gamma_{34}\frac{\rho_4}{\Gamma_0}=\Gamma\Gamma_0(1-\beta\beta_0)\frac{\rho_4}{\Gamma_0}=\Gamma(1-\beta\beta_0)\rho_4
\ee
If the unshocked ejecta are cold (i.e., in their rest frame the energy density is $e_4=\rho''_4c^2$ and the pressure is zero), the jump conditions across the RS give:
\begin{eqnarray}
\label{beta3}
\beta_3=\frac{\beta_{34}}{3}=\frac{\beta_0-\beta}{3(1-\beta\beta_0)}
\end{eqnarray}

The relation between $dt'$ and $dR$ can be derived from Eq.~\ref{dtdr}. Similary to Eq.~\ref{beta3} we can write $\beta_2=\frac{\beta_{12}}{3}$. In our notation $\beta_{12}=\beta$ and $\beta_2=\frac{\beta_{FS}-\beta}{1-\beta_{FS}\beta}$. After some simple algebra we obtain: $dt'=3dR/(4\beta\Gamma c)$.

While the FS travels a distance $dR$, the mass of the unshocked ejecta which crosses the RS is:
\be
\frac{dM_{0,3}}{dR}=4\pi R^2_{RS}\rho_4 \frac{\beta_0-\beta}{\beta}
\ee
If the initial width of the shell in the progenitor frame is $\Delta_0=ct_d$ (where $t_d$ is the prompt duration in the progenitor frame) and the radius of the shell is $R_{shell}$, then the density $\rho_4$ of the unshocked ejecta, measured in the observer frame is:
\be
\rho_4=\frac{M_0}{4\pi R_{shell}^2\Delta_0}
\ee
Assuming $R\sim R_{RS}\sim R_{shell}$ we finally find:
\be
\frac{dM_{0,3}}{dR}= \frac{M_0}{ct_d}\frac{\beta_0-\beta}{\beta}
\ee

\vskip 0.2 cm
\noindent
{\it Derivation of $dE'_{ad,3}$ and $E'_{int,3}$} -- The equations given in Section \ref{ourmodel} to estimate adiabatic losses, radiative losses and internal energy are still valid and describe the evolution of the FS and region 2. To derive the internal energy content of region 3 (and how it changes due to radiative and adiabatic losses and to the mass flux of the unshocked ejecta entering the RS) we follow a very similar approach. For the ejecta we assume equal number densities of electrons and protons, and thus $\mu_{e}=m_e/m_p$, $\mu_{p}=1$ and $\rho_{4,p}=\rho_{4,e}m_p/m_e>>\rho_{4,e}$.
In general, the microphysical parameters \epse, \epsp\ and the radiative efficiency \epsrad\ of the RS can differ from those characterizing the FS. We introduce the subscript '3' to refer to parameters of region 3.
Following the same approach adopted for the shocked CBM, we derive the following equation for the Lorentz factor of the electrons shocked by the RS:
\be
\gamma_{acc,e,3}-1=\epsilon_{e,3}(\gamma_{34}-1)\frac{m_p}{m_e}
\ee
After being heated, the electrons radiate a fraction $\epsilon_{rad,3}$ of their energy. Their random Lorentz factor decreases to $\gamma_{rad,e,3}-1=(1-\epsilon_{rad,3})(\gamma_{acc,e,3}-1)$. The post shock random Lorentz factor of protons is instead given by: 
\be
\gamma_{acc,p,3}-1=\epsilon_{p,3}(\gamma_{34}-1)
\ee
Due to adiabatic losses, these post-shock Lorentz factors of protons and electrons decrease down to $\gamma_{ad,p,3}$ and $\gamma_{ad,e,3}$. The comoving volume of the shocked ejecta is $V'\sim \frac{R^3}{\Gamma}\frac{\beta_{34}}{\beta}$, which reduces, for relativistic blast waves (i.e., $\beta\sim1$), to $V'\sim\frac{R^3\beta_{34}}{\Gamma}$. 
The largest contribution to the internal energy and adiabatic losses from the shocked ejecta comes from the relativistic phase of the RS, i.e. when $\gamma_{34}>>1$. In this case, the comoving volume is well approximated by $V'\sim \frac{R^3}{\Gamma}$, and the equations \ref{eq:p_ad} and \ref{pad} derived in section \ref{ourmodel} can still be applied to the shocked ejecta and used to estimate its internal energy and the adiabatic losses: 
\be
\label{eq:int energy RS}
E'_{int,3}(R)\!&=&\!4 \pi c^2\!\!\! \int_0^R\!\!\! 
dr r^2 \frac{\beta_0-\beta}{\beta}
\times \nonumber \\  
&&\left\{\rho_{4,p}[\gamma_{ad,p,3}(R,r)\!-\!1]\!+\!\rho_{4,e} [\gamma_{ad,e,3}(R,r)\!-\!1] \right\}\!\!\! \nonumber
\ee

\be
\label{eq:ad losses RS}
\frac{dE'_{ad,3}(R)}{dR}&=&-4 \pi c^2 \left(\frac{1}{R}-\frac{1}{3}
\frac{d \log \Gamma}{dR}\right)\int_0^R dr r^{2} \frac{\beta_0-\beta}{\beta}
\times \nonumber \\  
&& \left\{\rho_{4,p} \frac{{p}^2_{ad,p,3}(R,r)}{{\gamma_{ad,p,3}}(R,r)}+\rho_{4,e}  
\frac{{p}^2_{ad,e,3}(R,r)}{{\gamma_{ad,e,3}}(R,r)}\right\}~.
\ee

\section{Pre-acceleration}
\label{preacceleration}
\citet{beloborodov02} studied the interaction between the prompt $\gamma$-ray radiation and the ambient medium and showed that pair loading and pre--acceleration of the CBM may produce noticeable effects both on the dynamics of the blast wave and on the observed afterglow emission. According to his model, when the pre--acceleration of the medium is not negligible, the ejecta initially move into a cavity until they reach a radius $R_{gap}$. At $R=R_{gap}$ the blast wave forms and starts to collect matter enriched by pairs and pre--accelerated.
In our model, developed in section \ref{ourmodel}, we consider the possibility that the CBM is enriched by pairs, but we assume that it is at rest in the progenitor frame. Instead, if the medium has a radial motion characterized by a bulk Lorentz factor $\Gamma_{CBM}$ (which can depend on the radius),  some of the equations previously proposed should be modified. First, energy conservation (Eq. \ref{eq:energy conservation}) now reads (assuming a cold CBM):
\be
d\left[\Gamma(M_0+m)c^2+\Gamma_{\it eff} E'_{int}\right]= \Gamma_{CBM} dm\,c^2+\Gamma_{\it eff} dE'_{rad}~.
\ee
where $m(R<R_{gap})=0$.
Second, the energy dissipated at the shock depends on the relative Lorentz factor between the blast wave and the mass $dm$, which is no longer $\Gamma$ but $\Gamma_{rel}=\Gamma\Gamma_{CBM}(1-\beta\beta_{CBM})$:
\be
dE'_{sh}=(\Gamma_{rel}-1) \,dm\,c^2
\ee 
The equations for the internal energy and for radiative and adiabatic losses can be derived as before, noting that the relative Lorentz factor $\Gamma_{rel}$ enters the definition of the post-shock Lorentz factor of electrons and protons and the definition of the comoving volume, which is given by $V'\sim\frac{R^3\Gamma_{CBM}}{\Gamma_{rel}}$.

A physical model that aims to estimate $R_{gap}$ and the properties of the accelerated medium (its pair loading, its bulk Lorentz factor, and their dependence on the radius) is described in \citet{beloborodov02}, where the observational impacts on the afterglow emission are also discussed. In general, the early emission is expected to be softer, both due to the pair loading and to the pre-acceleration of the CBM. Light curves and spectra of the pair radiation are discussed in \citet{beloborodov05}.

\section*{Acknowledgments}
LN thanks SISSA and INAF-Osservatorio Astronomico di Brera for the kind hospitality during the completion of this work. 
LS gratefully thanks the Osservatorio Astronomico di Brera -- Merate for 
hospitality and stimulating discussions. 
LN is grateful to Pawan Kumar for helpful discussions.
LS is supported by NASA through Einstein Postdoctoral Fellowship grant number 
PF1--120090 awarded by the Chandra X--ray Center, which is operated by the 
Smithsonian Astrophysical Observatory for NASA under contract NAS8--03060.
We acknowledge the 2011 PRIN-INAF grant for financial support. 

\bibliography{biblio.bib}

\label{lastpage}

\end{document}